\documentclass[amssymb, preprintnumbers, showpacs, showkeys,aps,prb,superscriptaddress,twocolumn]{revtex4-1}

\usepackage{pgfplots}
\usepackage{tikz}
\usetikzlibrary{positioning}

\usepackage{color} 
\usepackage{xcolor} 
\usepackage{hyperref}
\usepackage{slashed}
\usepackage{graphicx}
\usepackage{amsmath}
\usepackage{latexsym}
\usepackage{epstopdf}
\usepackage{amsmath}
\usepackage{enumitem}
\usepackage{amssymb,amsmath}
\usepackage{multirow}
\usepackage{mathtools}
\usepackage{bbm}
\usepackage{bm}
\usepackage{amsfonts}
\usepackage{amssymb}
\usepackage{calligra}
\usepackage{calrsfs}
\usepackage{makecell}
\usepackage[normalem]{ulem}
\usepackage[USenglish,american]{babel}

\expandafter\ifx\csname package@font\endcsname\relax\else
 \expandafter\expandafter
 \expandafter\usepackage
 \expandafter\expandafter
 \expandafter{\csname package@font\endcsname}%
\fi
\begin{document}

\title{Thermodynamics of blackbody radiation in nonlinear electrodynamics}
%\title{Blackbody Radiation and Thermodynamic Properties in Nonlinear Electrodynamics}
%\title{Quantum Electrodynamics Vacuum Polarization Corrections to the Blackbody Radiation Law}

\author{I. Soares}%
\email{winacio@cbpf.br}
\affiliation{Centro Brasileiro de Pesquisas Físicas, Rua Dr. Xavier Sigaud, 150, URCA, Rio de Janeiro CEP 22290-180, RJ, Brazil}%

\author{R. Turcati}%
\email{turcati@cbpf.br}
\affiliation{Centro Brasileiro de Pesquisas Físicas, Rua Dr. Xavier Sigaud, 150, URCA, Rio de Janeiro CEP 22290-180, RJ, Brazil}%

\author{S. B. Duarte}%
\email{sbd@cbpf.br}
\affiliation{Centro Brasileiro de Pesquisas Físicas, Rua Dr. Xavier Sigaud, 150, URCA, Rio de Janeiro CEP 22290-180, RJ, Brazil}%

\begin{abstract}

We study the blackbody properties and the thermodynamic equilibrium quantities of a photon gas in the framework of nonlinear electrodynamics. In this vein, we take into account the photon propagation in a uniform external magnetic field in the weak field approximation, where an angular anisotropic energy density distribution appears in the frequency spectrum. The particular case when the photon propagates perpendicular to the background magnetic field is also discussed, which allows us to probe the strong field regime. We then derive a modified blackbody spectral distribution and the Stefan-Boltzmann law in this situation. Considerations about Wien's displacement law and the Rayleigh-Jeans formula are contemplated as well. Deviations from the thermodynamic quantities at thermal equilibrium such as energy, pressure, entropy, and heat capacity densities are obtained from the Helmholtz free energy. As an application, we study three nonlinear electrodynamics, namely, the Euler-Heisenberg, the generalized Born-Infeld, and the logarithmic electrodynamics. Possible implications on stellar systems with strong magnetic fields such as magnetars are discussed.

\end{abstract}

\maketitle

%************************************************************************
\section{Introduction}\label{introduction}

Over the past few decades, there has been a growing interest in using nonlinear electrodynamics to probe physical processes in the regime of strong electromagnetic fields. These studies include investigations in the physics of high-intensity lasers \cite{Karbstein:2019oej,Marklund:2008gj,Battesti:2012hf,King:2015tba}, intense magnetic fields in compact astrophysical objects \cite{Guzman-Herrera:2020ffc,Kim:2022xum,MosqueraCuesta:2004em}, and radiation propagation inside some materials \cite{Baggioli:2016oju,Bi:2021maw}, among others\cite{Sorokin:2021tge}. 

As is well known, quantum electrodynamics (QED) describes with very high and accurate precision all the electromagnetic phenomena in both classical and quantum scales \cite{ParticleDataGroup:2022pth}. Nevertheless, the vacuum polarization induces small deviations from the standard results of QED, leading to the appearance of new phenomena such as birefringence, photon-photon scattering, vacuum dichroism, and photon acceleration, among others \cite{Battesti:2012hf}. It is important to remark that these effects become relevant when there exist electric and magnetic fields up to a critical value, $\varepsilon_{c}\approx{m_{e}^{2}c^{3}}/e\hbar\approx10^{18}V/m\approx10^{9}T$, in some region of the space, where $m_{e}$ is the electron rest mass \cite{Schwinger:1951nm}. 

The phenomenological features associated with the QED vacuum polarization are usually studied in the framework of nonlinear electrodynamics\cite{Boillat:1970gw,Bialynicka-Birula:1970nlh,Sorokin:2021tge}. In this sense, a straightforward manner to emulate vacuum polarization effects is by introducing external background fields in the standard theoretical models \cite{Schwinger:1951nm}. In this scenario, phenomena such as birefringence can be easily studied by describing electromagnetic waves propagating in empty space.

From the theoretical perspective, nonlinear electrodynamics have been extensively investigated in a wide range of areas such as gravity, cosmology, and condensed matter systems\cite{Soleng:1995kn,Bronnikov:2000vy,Ayon-Beato:2000mjt,Cataldo:2000we,Burinskii:2002pz,Hassaine:2007py,Pan:2011vi,Hendi:2012zz,Cembranos:2014hwa,Gaete:2013dta,Gaete:2014nda,Gaete:2015qda,Kruglov:2014hpa,Kruglov:2015kda,Kruglov:2016ezw,Liu:2019rib,Gullu:2020ant,Balakin:2021arf,Balakin:2021jby,Dehghani:2021fwb}. Nonlinear electrodynamics also appears as an important ingredient in some fundamental scenarios such as string and M theory \cite{Gibbons:2001gy,Bandos:2020hgy}. From the experimental point of view, in turn, the investigation of electromagnetic phenomena in the strong field regime is a straightforward manner to probe not only the properties of the QED in the nonperturbative regime but also the effects in quantum field theory in general. Several experimental efforts are currently in progress in order to probe nonlinear effects of the electromagnetic field, which include the measurement of light by light scattering in Pb+Pb collisions at the Large Hadron Collider\cite{ATLAS:2017fur}, the photon splitting in strong magnetic fields\cite{Akhmadaliev:2001ik}, and experiments with laser beams crossing magnetic fields\cite{Luiten:2004py}, among others\cite{Brodin:2001zz}. Indeed, deviations from QED are also to be inspected by some experiments underway, which include the Station of Extreme Light (SEL), Europe's Light Infrastructure (ELI Project), and the ExaWatt Center for Extreme Light Studies (XCELS). These recent developments in experimental physics, which probe some fundamental symmetries in physics, also encourage a new look at the possibility of physics beyond the Standard Model (SM) of particle physics and fundamental interactions.  

Effective field theories are vastly used to describe several phenomena at high energies \cite{Bolmont:2020aed,Bandos:2020jsw}. Here, we will explore the photon propagation in the presence of a background magnetic field and the consequences to the thermodynamics of blackbody radiation through the study of three nonlinear models, namely, Euler-Heisenberg electrodynamics, the generalized Born-Infeld theory, and the logarithmic Lagrangian.

The structure of this paper is organized as follows. In Sec. \ref{generalframework} we review the main features of gauge and Poincarè invariant nonlinear electrodynamics theories.  In Sec. \ref{photonpropagation}, the wave propagation in a background electromagnetic field is derived, while in Sec. \ref{MDR}, the modified dispersion relation is obtained. Aspects related to the blackbody spectral density and thermodynamic equilibrium properties of the system are discussed in Sec. \ref{Termo}. The implications on the Euler-Heisenberg, generalized Born-Infeld, and logarithmic electrodynamics are contemplated, respectively, in Secs. \ref{EH}, \ref{BI}, and (\ref{Log}). The regime of strong fields is studied in Sec. \ref{perpendicularpropagation} Some comments about the obtained results are discussed in Sec. \ref{someremarks}. Our final remarks and further perspectives can be found in Sec. \ref{conclusions}.

We shall adopt the Gaussian units unless otherwise specified. In our conventions, the signature of the Minkowski metric is $\left(+,-,-,-\right)$.

%************************************************************************
\section{General Framework}\label{generalframework}

In this section, we will give a brief review of the main features of nonlinear electrodynamics theories. To accomplish that, we will restrict our analysis to the class of gauge and Lorentz invariant Lagrangians $\mathcal{L}=\mathcal{L}\left(\mathcal{F},\mathcal{G}\right)$
%\begin{eqnarray}
%\mathcal{L}=\mathcal{L}\left(F,G\right)
%\end{eqnarray}
formed by the invariant bilinear forms
\begin{eqnarray}\label{bilinear1}
\mathcal{F}&\equiv&-\frac{1}{4}F_{\mu\nu}F^{\mu\nu}=\frac{1}{2}\left(\mathbf{E}^{2}-\mathbf{B}^{2}\right),\\
\label{bilinear2}\mathcal{G}&\equiv&-\frac{1}{4}F_{\mu\nu}\tilde{F}^{\mu\nu}=\mathbf{E}\cdot\mathbf{B},
\end{eqnarray}
where $F_{\mu\nu}\left(\equiv\partial_{\mu}A_{\nu}-\partial_{\nu}A_{\mu}\right)$ is the field strength of the electromagnetic field and $\tilde{F}^{\mu\nu}=\left(1/2\right)\epsilon^{\mu\nu\alpha\beta}F_{\alpha\beta}$ is the dual stress tensor. To preserve the parity symmetry, only quadratic terms in the fields will be considered. 

The full description of the system consists of the dynamical equation for the electromagnetic field 
%\begin{eqnarray}
%\mathcal{L}=\mathcal{L}\left(\mathcal{F},\mathcal{G}\right)
%\end{eqnarray}
\begin{eqnarray}\label{fieldequation}
%\partial_{\mu}h^{\mu\nu}=0,
\partial_{\nu}\left(\frac{\partial\mathcal{L}}{\partial{F_{\mu\nu}}}\right)=0,
\end{eqnarray}
plus the Bianchi identity
\begin{eqnarray}\label{Bianchi}
%\partial_{\mu}\tilde{h}^{\mu\nu}=0,
\partial_{\alpha}F_{\mu\nu}+\partial_{\mu}F_{\nu\alpha}+\partial_{\nu}F_{\alpha\mu}=0.
\end{eqnarray}

Taking the invariant bilinear forms (\ref{bilinear1}) and (\ref{bilinear2}) into account, the field equation for the parity-conserving nonlinear theory takes the following form:
\begin{eqnarray}\label{equationofmotion}
c_{1}\partial_{\nu}F^{\mu\nu}-\frac{1}{2}M^{\mu\nu\alpha\beta}\partial_{\nu}F_{\alpha\beta}=0, 
\end{eqnarray}
where 
\begin{eqnarray}\label{generaltensor}
M^{\mu\nu\alpha\beta}&=&d_{1}F^{\mu\nu}F^{\alpha\beta}+d_{2}\tilde{F}^{\mu\nu}\tilde{F}^{\alpha\beta}\nonumber\\
&&+d_{3}\left(F^{\mu\nu}\tilde{F}^{\alpha\beta}+\tilde{F}^{\mu\nu}F^{\alpha\beta}\right)+c_{2}\epsilon^{\mu\nu\alpha\beta}, 
\end{eqnarray}
and
\begin{eqnarray}\label{coefficients}
c_{1}&=&\frac{\partial\mathcal{L}}{\partial\mathcal{F}}\bigg|_{\mathbf{E,B}}, \quad 
c_{2}=\frac{\partial\mathcal{L}}{\partial\mathcal{G}}\bigg|_{\mathbf{E,B}}, \quad
d_{1}=\frac{\partial^{2}\mathcal{L}}{\partial\mathcal{F}^{2}}\bigg|_{\mathbf{E,B}},\nonumber\\
&&d_{2}=\frac{\partial^{2}\mathcal{L}}{\partial\mathcal{G}^{2}}\bigg|_{\mathbf{E,B}}, \quad
d_{3}=\frac{\partial^{2}\mathcal{L}}{\partial\mathcal{F}\partial\mathcal{G}}\bigg|_{\mathbf{E,B}}.
\end{eqnarray}

The tensor $M^{\mu\nu\alpha\beta}$ is symmetric with respect to
the exchange of the pairs of indices $\mu\nu$ and $\alpha\beta$, and antisymmetric with respect to the exchange of indices within each pair. In addition, when one inserts the tensor $M^{\mu\nu\alpha\beta}$ into the equation of motion, the Levi-Civita tensor contribution drops out because of the
Bianchi's identity, while the remaining pieces reproduce the photon dynamical equation in the framework of nonlinear electromagnetism. Note also that the coefficients $c_{1}$, $c_{2}$, $d_{1}$, $d_{2}$, and $d_{3}$ are all evaluated at the external fields $\mathbf{E}$ and $\mathbf{B}$.

%*******************************************
\subsection{Photon propagation in an external electromagnetic field}\label{photonpropagation}

The photon propagation in an external electromagnetic field will be described as weak field disturbances propagating around this background field. At this level, the equation of motion for the electromagnetic wave is linear, and the influence of the external field will be encoded in the coefficients in (\ref{coefficients}).

We now pass on the calculation of the field equation for the photon in the present scenario. We start by adopting the linearization procedure and splitting the electromagnetic field $F^{\mu\nu}$ as
\begin{equation}\label{split}
F^{\mu\nu}=F^{\mu\nu}_{B}+f^{\mu\nu},
\end{equation}
where $F^{\mu\nu}_{B}$ describes a classical background electromagnetic field and $f^{\mu\nu}$ is a perturbation wave field.  Inserting the relation (\ref{split}) into Eq. (\ref{fieldequation}), and assuming that the background field satisfies the field equations, one finds 
\begin{eqnarray}\label{linearizedequation}
\partial_{\nu}\left(\Omega^{\mu\nu\alpha\beta}f_{\alpha\beta}\right)=0, 
\end{eqnarray}
where
\begin{eqnarray}
\Omega^{\mu\nu\alpha\beta}=\frac{\partial^{2}\mathcal{L}}{\partial{F_{\mu\nu}}\partial{F_{\alpha\beta}}}\bigg|_{B}. 
\end{eqnarray}

The above tensor holds the same symmetries as the tensor $M^{\mu\nu\alpha\beta}$, and the subscript $B$ means that $\Omega^{\mu\nu\alpha\beta}$ is evaluated at the background electromagnetic fields.

Next, considering the invariant bilinear forms (\ref{bilinear1}) and (\ref{bilinear2}), the field equations associated to the perturbation field $f_{\mu\nu}$ are 
\begin{eqnarray}
c_{1}\partial_{\nu}f^{\mu\nu}-\frac{1}{2}M^{\mu\nu\alpha\beta}_{B}\partial_{\nu}f_{\alpha\beta}=0. 
\end{eqnarray}

We are now bound to consider the regime of slow varying but arbitrary background electromagnetic fields. In this context, and considering the decomposition in Fourier modes of the field $f_{\mu\nu}$, Eq. (\ref{linearizedequation}) takes the form
\begin{eqnarray}\label{omega}
\Omega^{\mu\nu\alpha\beta}k_{\nu}f_{\alpha\beta}=0. 
\end{eqnarray}

Furthermore, the Bianchi identity now reads as
\begin{eqnarray}
\partial_{\alpha}f_{\mu\nu}+\partial_{\mu}f_{\nu\alpha}+\partial_{\nu}f_{\alpha\mu}=0,
\end{eqnarray}
which restricts the wave field $f^{\mu\nu}$ to be of the form
\begin{eqnarray}
f_{\mu\nu}=\partial_{\mu}a_{\nu}-\partial_{\nu}a_{\mu}, 
\end{eqnarray}
where $a^{\mu}$ is the gauge field associated to the stress tensor $f_{\mu\nu}$.

In terms of the gauge field $a^{\mu}$, Eq. (\ref{omega}) yields
\begin{eqnarray}
\Omega^{\mu\nu\alpha\beta}k_{\nu}k_{\beta}a_{\alpha}=0,
\end{eqnarray}
where the tensorial quantity $\Omega^{\mu\nu\alpha\beta}$ can be written as
\begin{eqnarray}\label{tensor}
\Omega^{\mu\nu\alpha\beta}=c_{1}\left(\eta^{\mu\alpha}\eta^{\nu\beta}-\eta^{\mu\beta}\eta^{\nu\alpha}\right)-M^{\mu\nu\alpha\beta}_{B}, 
\end{eqnarray}
which contains an isotropic part plus an anisotropic contribution $M^{\mu\nu\alpha\beta}_{B}$, which comes from the nonlinearity of the electromagnetic field.

%***********************************************
\subsection{Modified dispersion relation}\label{MDR}

As discussed in the previous section, the equation of motion associated with the weak disturbance is linear, where the coefficients depend on the external background field. To get a better understanding of wave propagation in this situation, we will derive the dispersion relation for the electromagnetic wave in the presence of background magnetic fields.

Let us then start off our considerations by taking into account the standard procedure to find the electromagnetic wave frequencies, which consists of solving the system of linear equations
\begin{eqnarray}
A^{\mu\alpha}\epsilon_{\alpha}=0, 
\end{eqnarray}
where the tensor $A^{\mu\alpha}$ is defined to be
\begin{eqnarray}
A^{\mu\alpha}=\Omega^{\mu\nu\alpha\beta}k_{\nu}k_{\beta}, 
\end{eqnarray}
and we have defined the normalized polarization tensor $\epsilon_{\mu}=a_{\mu}/\sqrt{a^{2}}$.

According to definition (\ref{tensor}), the above tensor can be cast under the form
\begin{eqnarray}
A^{\mu\alpha}\equiv{c}_{1}\left(\eta^{\mu\alpha}k^{2}-k^{\mu}k^{\alpha}\right)-M^{\mu\nu\alpha\beta}k_{\nu}k_{\beta}.  
\end{eqnarray}

The corresponding theory is gauge invariant, which means that there exist spurious modes, and a gauge fixing becomes necessary. One possible choice is to adopt the temporal gauge $a^{0}=0$. This choice has the advantage that immediately removes 1 degree of freedom from the gauge field $a^{\mu}$. 

In what follows, we then adopt the temporal gauge, which decomposes the system of linear equations in
\begin{eqnarray}\label{temporalequation}
A^{0i}\epsilon_{i}=0 
\end{eqnarray}
and the reduced system
\begin{eqnarray}\label{spatialequation}
A^{ij}\epsilon_{j}=0.
\end{eqnarray}

Our main purpose in this work is to find the blackbody radiation laws in the presence of a background magnetic field. Therefore, it will be considered an external uniform magnetic field $\mathbf{B}$, where the electric field will be neglected, i.e., $\mathbf{E}=\mathbf{0}$. Note that only the coefficients $c_{1}$, $d_{1}$, and $d_{2}$ are nonzero in this configuration. 

With these assumptions, and assuming $k^{\mu}=\left(w/c,\mathbf{k}\right)$, Eq. (\ref{temporalequation}) provides us with
\begin{eqnarray}
\mathbf{k}\cdot\mathbf{\epsilon}=-\frac{d_{2}}{c_{1}}\left(\mathbf{k}\cdot\mathbf{B}\right)\left(\mathbf{B}\cdot\mathbf{\epsilon}\right). 
\end{eqnarray}

Next, taking into account the above relation, Eq. (\ref{spatialequation}) takes the form
\begin{eqnarray}\label{matrixequation}
&&\left[\left(\frac{w^{2}}{c^{2}}-\mathbf{k}^{2}\right)\delta_{ij}+\frac{d_{1}}{c_{1}}(\mathbf{k}\times\mathbf{B})_{i}(\mathbf{k}\times\mathbf{B})_{j}\right.\nonumber\\
&&\left.-\frac{d_{2}}{c_{1}}(\mathbf{k}\cdot\mathbf{B})k_{i}B_{j}+\frac{d_{2}}{c_{1}}\frac{w^{2}}{c^{2}}{B}_{i}{B}_{j}\right]\epsilon_{j}=0. 
\end{eqnarray}

A necessary and sufficient condition for the eigenvalue problem above to have solutions different from the trivial one is to find the vanishing determinant of the matrix (\ref{matrixequation}). An explicit calculation gives us
\begin{eqnarray}
detA^{ij}=\left(\frac{w^{2}}{c^{2}}-\mathbf{k}^{2}\right)\mathcal{P}_{4}\left(k\right).    
\end{eqnarray}

The determinant is a sixth-order polynomial. However, the physically relevant part is given by the fourth-order polynomial $\mathcal{P}_{4}\left(k\right)$ in the variables $w$ and $k$, which is explicitly given by
\begin{eqnarray}\label{fourthpoly}
\mathcal{P}_{4}\left(k\right)=Pw^{4}+Qw^{2}+R,\end{eqnarray}
where
\begin{eqnarray}
P&=&\frac{1}{c^{4}}\left(1+\frac{d_{2}}{c_{1}}\mathbf{B}^{2}\right),\\
Q&=&\frac{1}{c^{2}}\left[-2\mathbf{k}^{2}+\frac{d_{1}}{c_{1}}\left(\mathbf{k}\times\mathbf{B}\right)^{2}-\frac{d_{2}}{c_{1}}\left[\left(\mathbf{k}\cdot\mathbf{B}\right)^{2}+\mathbf{k}^{2}\mathbf{B}^{2}\right]\right.\nonumber\\
&&\left.+\frac{d_{1}d_{2}}{c_{1}^{2}}\left(\mathbf{k}\times\mathbf{B}\right)^{2}\mathbf{B}^{2}\right],\\
R&=&\mathbf{k}^{4}-\frac{d_{1}}{c_{1}}\mathbf{k}^{2}\left(\mathbf{k}\times\mathbf{B}\right)^{2}+\frac{d_{2}}{c_{1}}\mathbf{k}^{2}\left(\mathbf{k}\cdot\mathbf{B}\right)^{2}\nonumber\\
&&-\frac{d_{1}d_{2}}{c_{1}^{2}}\left(\mathbf{k}\cdot\mathbf{B}\right)^{2}\left(\mathbf{k}\times\mathbf{B}\right)^{2}.
\end{eqnarray}

The dynamical polarization states are given by linearly independent solutions of the eigenvalue problem (\ref{spatialequation}) under the nontrivial solutions of the condition $detA^{ij}=0$, which, according to (\ref{fourthpoly}), provide us with four solutions. On the other hand, because of the $CPT$ invariance, if $k=\left(-w,\mathbf{k}\right)$ is a solution, then $-k=\left(w,-\mathbf{k}\right)$ is a solution as well. Therefore, we have a two-dimensional space of polarization states. 

The wave frequencies, therefore, take the following form:
\begin{eqnarray}\label{frequency1}
w_{1}\left(\mathbf{k}\right)&=&ck\sqrt{1-\frac{d_{1}}{c_{1}}\left(\mathbf{\hat{k}}\times\mathbf{B}\right)^{2}}, \\
\label{frequency2}w_{2}\left(\mathbf{k}\right)&=&ck\sqrt{1-\frac{d_{2}\left(\mathbf{\hat{k}}\times\mathbf{B}\right)^{2}}{c_{1}+d_{2}\mathbf{B}^{2}}}. 
\end{eqnarray}

The frequencies above are associated with wave propagation in a magnetized medium from the nonlinear electrodynamics perspective. Furthermore, these dispersion relations are distinct, which leads to the phenomenon of birefringence\cite{Bialynicka-Birula:1970nlh}. We also remark that $d_{1}\rightarrow0$ and $d_{2}\rightarrow0$, or, equivalently, whenever $\mathbf{B}\rightarrow{0}$, the standard photon frequencies are recovered. The conditions $c_{1}>{d_{1}}\left(\mathbf{\hat{k}}\times\mathbf{B}\right)^{2}$ and $c_{1}+{d_{2}}\left(\mathbf{\hat{k}}\cdot\mathbf{B}\right)^{2}>0$ ensure that frequencies (\ref{frequency1}) and (\ref{frequency2}) are real and positive definite.

The corresponding frequencies can be written in terms of the angle $\theta$ between the wave vector $\mathbf{k}$ and the external magnetic field $\mathbf{B}$, which gives us
\begin{eqnarray}
w_{1}\left(k\right)&=&ck\sqrt{1-\frac{d_{1}}{c_{1}}B^{2}sin^{2}\theta}, \\
w_{2}\left(k\right)&=&ck\sqrt{1-\frac{d_{2}B^{2}}{c_{1}+d_{2}{B}^{2}}sin^{2}\theta}. 
\end{eqnarray}

Here one notes that whenever the wave vector $\mathbf{k}$ and the background magnetic field $\mathbf{B}$ are perpendicular to each other, the frequencies reduce to $w_{1}\left({k}\right)=ck\left(1-d_{1}B^{2}/c_{1}\right)^{1/2}$ and %$w_{2}\left({k}\right)=ck\sqrt{1-d_{2}B^{2}/\left(c_{1}+d_{2}B^{2}\right)}$ 
$w_{2}\left({k}\right)=ck\left(1+d_{2}B^{2}/c_{1}\right)^{-1/2}$, respectively.  

The group velocities, in turn, related to the above frequencies, are given by\cite{Neves:2021tbt}
\begin{eqnarray}\label{groupvelocity}
\mathbf{v}_{g}^{(1)}&=&c\frac{\left[c_{1}\mathbf{\hat{k}}-d_{1}\mathbf{B}\times\left(\mathbf{\hat{k}}\times\mathbf{B}\right)\right]}{c_{1}\sqrt{1-\frac{d_{1}}{c_{1}}\left(\mathbf{\hat{k}}\times\mathbf{B}\right)^{2}}},\\
\mathbf{v}_{g}^{(2)}&=&c\frac{\left[c_{1}\mathbf{\hat{k}}+d_{2}\mathbf{B}\left(\mathbf{\hat{k}}\cdot\mathbf{B}\right)\right]}{\left(c_{1}+d_{2}\mathbf{B}^{2}\right)\sqrt{1-\frac{d_{2}\left(\mathbf{\hat{k}}\times\mathbf{B}\right)^{2}}{c_{1}+d_{2}\mathbf{B}^{2}}}},
\end{eqnarray}
which have components in the directions of $\mathbf{\hat{k}}$ and $\mathbf{B}$. Furthermore, whenever $B\rightarrow0$, one recovers the Maxwell theory, and the group velocity goes to $\mathbf{v}_{g}=\mathbf{\hat{k}}w/k$.

We would like to stress that the nonlinear features of the quantum vacuum were treated as a classical medium. An alternative description implies considering the modifications in the vacuum as an effective geometry for the photon propagation\cite{Boillat:1970gw,Bialynicka-Birula:1970nlh}. In such a case, photons propagate as null geodesics in a background metric that deviates from the Minkowski one due to the nonlinearities of the electromagnetic field. Although these formalisms describe distinct situations, both approaches are described in the soft photon approximation and provide exactly the same results for the frequency modes. At this point, it is perhaps worth remarking that there exist three different situations where the effective metric emerges in nonlinear electrodynamics, which are due to G. Boillat\cite{Boillat:1970gw}, Bialynicka-Birula and Bialynicki-Birula\cite{Bialynicka-Birula:1970nlh}, and Novello and coworkers\cite{Novello:1999pg}. The effective metrics in Boillat and Bialynicka-Birula and Bialynicki-Birula are entirely equivalent, while the Novello effective metric is only conformally equivalent to the mentioned ones\cite{deMelo:2014isa}. The difference between both approaches is due to the schematic procedure adopted by the authors to obtain the effective geometry. While Boillat-Birula frameworks take the eikonal approximation into account, Novelo geometry is derived considering the Hadamard theory, which provides distinct coefficients when compared with the Boillat-Birula metrics. However, these effective metrics can be connected through a conformal factor, which can vanish in some special cases\cite{deMelo:2014isa}. 

To conclude this section, we would like to emphasize that our approach is equivalent to the Boillat-Birula frameworks. Nevertheless, different from these works, here we have explored the gauge structure of the system to get the wave frequencies of the nonlinear electromagnetic wave\cite{Liberati:2000mp}.

%*************************************************************
\subsection{Blackbody radiation and thermodynamic properties of the photon gas}\label{Termo}

Our goal in this section is to use the techniques of statistical mechanics to derive the frequency spectrum and the thermodynamic quantities of a photon gas in the framework of nonlinear electrodynamics. The fundamental object for this analysis is the partition function $\mathcal{Z}$. In our approach, it will be considered nonzero temperatures below the electron rest mass $m_{e}$, i.e., $k_{B}T\ll{m_{e}c^{2}}$, which will enable us to use the effective field theory to compute the free energy of the photon field. Indeed, at temperatures well below the electron rest mass, the electron-positron concentration is exponentially small, i.e., proportional to $exp\left(-m_{e}c^{2}/k_{B}T\right)$, and the contributions to the thermodynamic properties of the blackbody radiation mainly come from the photon sector \cite{Barton:1990mu,Kong:1998ic}. Furthermore, the partition function will be formulated in the grand canonical potential for the photon gas  with zero chemical potential assuming the Bose-Einstein statistics \cite{NiauAkmansoy:2013sxs,Anacleto:2018wlj}.

%**************************************
\subsubsection{The partition function and the spectral energy density}

As stated above, we need to find the partition function in order to derive the blackbody radiation and the thermodynamic quantities at thermal equilibrium. To begin with, one notes that the number of available states $N$ for a given system is
\begin{eqnarray}
N=\int{d\mathbf{x}}\int\frac{{d\mathbf{k}}}{\left(2\pi\right)^{3}}.
\end{eqnarray}

In spherical coordinates, the above equation can be written as
\begin{eqnarray}
N=\frac{V}{\left(2\pi\right)^{3}}\int{d}\Omega\int_{0}^{\infty}{dkk^{2}},
\end{eqnarray}
where $V$ is the volume of the reservoir and $d\Omega$ is the solid-angle element.  

To find the number of states $N$ for which the photon frequency lies between $\nu$ and $\nu+d\nu$, one needs to transform the above $k$ integral to the frequency $\nu$ space. To achieve that, it is necessary to take into account both the phase $v_{p}$ and group $v_{g}$ velocities previously derived. However, in the general case, one gets a very complicated integral. To circumvent this problem, we consider the weak field approximation, which is obtained by imposing the following conditions: $c_{1}\gg{d_{1}}\left(\mathbf{\hat{k}}\times\mathbf{B}\right)^{2}$ and $c_{1}\gg-{d_{2}}\left(\mathbf{\hat{k}}\cdot\mathbf{B}\right)^{2}$. This condition means that our approach will be restricted to situations in which one has small deviations from the Maxwell theory. In this regime, the modulus of the phase and the group velocities for each mode are equal and given, respectively, by\cite{Bialynicka-Birula:1970nlh}
\begin{eqnarray}
v_{p}^{(1)}=v_{g}^{(1)}=c\left[1-\frac{d_{1}}{2c_{1}}\left(\mathbf{\hat{k}}\times\mathbf{B}\right)^{2}\right],
\end{eqnarray}
and
\begin{eqnarray}
v_{p}^{(2)}=v_{g}^{(2)}=c\left[1-\frac{d_{1}\left(\mathbf{\hat{k}}\times\mathbf{B}\right)^{2}}{2\left(c_{1}+d_{2}\mathbf{B}^{2}\right)}\right].
\end{eqnarray}

Therefore, if one substitutes ${k}^{2}$ by the dispersion relations (\ref{frequency1}) and (\ref{frequency2}), in the weak field approximation, one promptly gets
\begin{eqnarray}
dk_{1,2}=\frac{2\pi}{c}\frac{d\nu}{\Lambda_{1,2}}
\end{eqnarray}
for each mode, where $\Lambda_{1,2}$ are defined as
\begin{eqnarray}\label{lambda1}
\Lambda_{1}&=&1-\frac{d_{1}}{2c_{1}}B^{2}sin^{2}\theta, \\
\label{lambda2}\Lambda_{2}&=&1-\frac{d_{2}B^{2}sin^{2}\theta}{2\left(c_{1}+d_{2}{B}^{2}\right)}. 
\end{eqnarray}

Hence, the number of available states $N$ reads as %assume the form
\begin{eqnarray}\label{Nstates}
N=N_{1}+N_{2}=\frac{V}{c^{3}}\int{d\Omega}\int_{0}^{\infty}d\nu\nu^{2}\Delta\Lambda\left(B,\theta\right),
\end{eqnarray}
where $\Delta\Lambda\left(B,\theta\right)$ is given by
\begin{eqnarray}\label{factororiginal}
\Delta\Lambda\left(B,\theta\right)\equiv\left(\frac{1}{\Lambda_{1}^{3}}+\frac{1}{\Lambda_{2}^{3}}\right)\approx2+\epsilon{sin^{2}\theta},
\end{eqnarray}
and
\begin{eqnarray}
\epsilon=\frac{3d_{1}B^{2}}{2c_{1}}\left[1+\frac{\left(d_{2}/d_{1}\right)}{1+\left(d_{2}/c_{1}\right)B^{2}}\right].\end{eqnarray}

Note that $\Delta\Lambda\left(B,\theta\right)$ depends on the magnitude of the background magnetic field $B$ and the $\theta$ angle between the wave vector $\mathbf{k}$ and the magnetic field $\mathbf{B}$. In the special case when the photon propagation is perpendicular to the magnetic field, the factor $\Delta\Omega\left(B\right)$ depends only on the magnitude of the magnetic field. Furthermore, whenever $B\rightarrow0$, $\Delta\Lambda=2$, and the number of available states of a photon gas in the Maxwell theory are recovered. We also remark that when the photon propagation is parallel to the background magnetic field $\left(\theta=0\right)$, $\Delta\Lambda=2$, and the photon propagates at the speed of light.

Having characterized the regime of validity of our formalism, we are now ready to obtain the partition function $\mathcal{Z}$ in this situation. Following the standard methodology, the logarithm of the partition function $\mathcal{Z}$ reads as
\begin{eqnarray}\label{logZ}
log\mathcal{Z}&=&-\frac{V}{c^{3}}\int {d\Omega}\times\nonumber\\
&&\int^{\infty}_{0}{d}\nu{\nu^{2}}\Delta\Lambda\left(B,\theta\right)log\left(1-e^{-\beta{h\nu}}\right).
\end{eqnarray}

From relation (\ref{logZ}), one can derive the frequency spectrum and the related thermodynamic quantities. 

The spectral energy density $u$, per unit volume, in thermal equilibrium at temperature $T$ is then given by
\begin{eqnarray}\label{spectralenergydensity}
u\left(\nu,T\right)=\left(\frac{8\pi\nu^{2}}{c^{3}}\right)\left(1+\frac{\epsilon}{3}\right)\frac{h\nu}{\left(e^{\beta{h}\nu}-1\right)}.
\end{eqnarray}

A quick glance at the energy density (\ref{spectralenergydensity}) clearly shows us that the contribution from the nonlinearities is encoded in the $\epsilon$ parameter. In the limit $\epsilon=0$, i.e., whenever $B\rightarrow0$, or, equivalently, $d_{1}\rightarrow{0}$ and $d_{2}\rightarrow{0}$, the internal energy density $u\left(\nu,T\right)$ reduces to the Planck distribution at the temperature T, as expected. Furthermore, the number $3$ in $\epsilon/3$ has a geometric origin since it arrives from the angular integration of the factor $\Delta\Lambda\left(B,\theta\right)$ in (\ref{logZ}).

At low frequencies, the frequency distribution (\ref{spectralenergydensity}) assumes the form
\begin{eqnarray}
u\left(\nu,T\right)=\left(\frac{8\pi\nu^{2}}{c^{3}}\right)\left(1+\frac{\epsilon}{3}\right)\left(k_{B}T\right).
\end{eqnarray}

From the above relation, we arrive at the conclusion that the Rayleigh-Jeans law is modified due to a background magnetic field. On the other hand, Wien's displacement law is not changed in this context.

Integrating (\ref{spectralenergydensity}) over all the frequencies, the total energy density obtained is 
\begin{eqnarray}\label{totalenergydensity}
u\left(T\right)=aT^{4}, 
\end{eqnarray}
with 
\begin{eqnarray}\label{lettera}
a=\frac{4}{c}\left(\frac{2\pi^{5}k_{B}^{4}}{15h^{3}c^{2}}\right)\left(1+\frac{\epsilon}{3}\right)
\end{eqnarray}
being an effective coefficient that retains the nonlinear modifications. 

With regard to the angular dependence, the energy density contribution for each solid-angle element is given by
\begin{eqnarray}
u\left(T,\Omega\right)d\Omega =\left(\frac{2\pi^{4}\kappa_{B}^{4}}{15h^{3}c^{3}}\right)T^{4}\left(1+\frac{\epsilon}{2}sin^{2}\theta\right)d\Omega.
\end{eqnarray}

Thus, the angular energy distribution induces the appearance of a quadrupole $\left(l=2\right)$ term to the power angular spectrum, which gives an anisotropic contribution to the frequency spectrum. Furthermore, we could have expanded the factor (\ref{factororiginal}) at higher orders in the binomial approximation, which would give additional contributions to the power angular spectrum of the order $l=2n$. To achieve this, we would have to impose new constraints on the magnitude of the magnetic field $B$. This result can play an important role in the anisotropies of the cosmic microwave background. 

%****************************
\subsubsection{Radiance and the modified Stefan-Boltzmann law}

The radiance is defined by the total energy emitted per unit of time and per unit area of the cavity surface. For a photon gas in the Maxwell theory, the spectral radiance $B\left(\nu,T\right)$ emitted from the blackbody surface is isotropic and depends only on the frequency $\nu$ and the temperature $T$. Here, on the other hand, the spectral radiance \begin{eqnarray}\label{intensityoflight}
B\left(\nu,\theta,T\right)=\frac{\nu^{2}}{c^{2}}\left(\frac{h\nu}{e^{\beta{h}\nu}-1}\right)\Delta\Lambda\left(B,\theta\right),    
\end{eqnarray}
depends also on the background magnetic field $B$ and the angle $\theta$ between the wave vector $\mathbf{k}$ and the external magnetic field.

Regarding the radiance, the explicit form can be found by solving the following integral
\begin{eqnarray}\label{definition}
R\left(T\right)=\int_{0}^{2\pi}{d\phi}\int_{0}^{\pi/2}d\theta{sin\theta}cos\theta\int_{0}^{\infty}B\left(\nu,\theta,T\right)d\nu.   
\end{eqnarray}

By means of this integral, it is straightforward to find the Stefan-Boltzmann law, which provide us with
\begin{eqnarray}\label{stefanboltzmannlaw}
R\left(T\right)=\sigma_{eff}T^{4},
\end{eqnarray}
where 
\begin{eqnarray}\label{effectivesigma}
\sigma_{eff}=\sigma\left(1+\frac{\epsilon}{4}\right),
\end{eqnarray}
is the effective Stefan-Boltzman constant and $\sigma=\left(2\pi^{5}k_{B}^{4}/15h^{3}c^{2}\right)$ is the usual Stefan-Boltzman constant. The factor $\epsilon$ carries the nonlinear modifications in the Stefan-Boltzmann constant.

As is well known, the radiance and the energy density are proportional to each other, being related by purely geometric factors. On the other hand, in the present context, this relation is not preserved anymore, and there appears to be a dependence on the specific nonlinear model, as one can see by evaluating the relations (\ref{lettera}) and (\ref{effectivesigma}). Indeed, the emergence of angular dependence in the spectral radiance (\ref{intensityoflight}) changes the solid-angle integral of the radiance (\ref{definition}), and the connection between the energy density and the radiance through geometric factors is lost. More specifically, it is the appearance of the quadrupole moment in the frequency spectrum induced by the nonlinearity that breaks the relation between the mentioned quantities. In the particular case where the wave propagation is perpendicular to the external magnetic field, there is no dependence on the angle (while there still exists the influence of the magnetic field), and the connection between the radiance and the energy density through the factor $c/4$ is recovered. This feature will be commented on in Sec. \ref{perpendicularpropagation}.

%*******************************************

\subsubsection{Thermodynamic quantities}

We can further investigate the consequences of the nonlinearity in the photon gas sector by evaluating the thermodynamic variables. In this sense, we first obtain the free energy $F$, namely, %which takes the form
\begin{eqnarray}\label{freeenergy}
F=-V\left(\frac{8\pi^{5}k_{B}^{4}{T}^{4}}{45h^{3}c^{3}}\right)\left(1+\frac{\epsilon}{3}\right).
\end{eqnarray}

The pressure $p$, the energy $u$, the entropy $s$, and the heat capacity $c_{V}$ at constant volume densities are, respectively, given by
\begin{eqnarray}\label{pressure}
p=\frac{a}{3}T^{4}, \quad u=aT^{4}, \quad
s=\frac{4}{3}aT^{3}, 
\end{eqnarray}
and
\begin{eqnarray}\label{heat}
c_{V}=4aT^{3},
\end{eqnarray}
with $a$ being defined on relation (\ref{lettera}).

Relations (\ref{freeenergy}), (\ref{pressure}), and (\ref{heat}) show us that the electromagnetic wave propagation in a magnetized medium modifies these quantities, leading to deviations of the free energy and the corresponding derived thermodynamic equilibrium quantities. On the other hand, the equation of state that relates energy and pressure densities is maintained even in the presence of background magnetic fields, i.e., $p=u/3$.

%************************************************

\section{Application to Nonlinear Electrodynamics Models}\label{application}

We now apply the above framework to three nonlinear electrodynamic models: the Euler-Heisenberg, the generalized Born-Infeld, and the logarithmic electrodynamics. In addition, we will explore the wave propagation perpendicular to the external magnetic field. In this case, the angular dependence vanishes and we can compute the integral
(\ref{logZ}) without any approximation, which allows us to study the regime of strong magnetic fields for the Born-Infeld and the logarithmic electrodynamics.

%*********************************************

\subsection{The Euler-Heisenberg effective Lagrangian}\label{EH}

The Euler-Heisenberg theory is a full nonperturbative effective action that describes the quantum electrodynamics vacuum polarization effects at one loop order in the presence of a uniform background electromagnetic field \cite{Heisenberg:1936nmg,Dunne:2012vv}. These effects become relevant above the critical field $\mathcal{E}_{c}$, the so-called Schwinger limit, where there is the production of real electron-positron pairs. 

The density Lagrangian of the aforementioned model is given by %(hep-th/0406216 - sistema gaussiano)
\begin{eqnarray}\label{LEH}
&&\mathcal{L}_{EH}=\mathcal{F}-\frac{1}{8\pi^{2}}\int_{0}^{\infty}\frac{ds}{s^{3}}e^{-m^{2}s}\nonumber\\
&&\times\left[\left(es\right)^{2}\mathcal{G}\frac{\mathcal{R}cosh\left(es\sqrt{-\mathcal{F}+i\mathcal{G}}\right)}{\mathcal{I}cosh\left(es\sqrt{-\mathcal{F}+i\mathcal{G}}\right)}+\frac{2}{3}\left(es\right)^{2}\mathcal{F}-1\right],\nonumber\\
\end{eqnarray}
where $\mathcal{R}$ and $\mathcal{I}$ are related to the real and imaginary parts, respectively.

In the weak field limit of the Euler-Heisenberg electrodynamics, the Lagrangian density reduces to \cite{Adler:1970gg,Ritus:1972ky}
\begin{eqnarray}
\mathcal{L}_{EH}=\mathcal{F}+\frac{2\alpha^{2}\hbar^{3}}{45m_{e}^{4}c^{5}}\left(4\mathcal{F}^{2}+7\mathcal{G}^{2}\right),
\end{eqnarray}
where $\alpha=e^{2}/\hbar{c}$. 

The weak field limit of the Euler-Heisenberg electrodynamics is justified if the dimensionless expansion parameter $4\pi\alpha\hbar^{3}|F|^{2}/\left(m_{e}^{4}c^{4}\right)$ is much smaller than unity\cite{Bialynicka-Birula:1970nlh}. This is the case, for instance, for strong magnetic fields in neutron stars that may be as large as
$10^{12}$Gauss\cite{Gold:1968zf}, where processes such as photon splitting and pair
conversion are expected to occur in the vicinity of these compact objects\cite{Baring:2008aw}.

In accordance with our formalism previously developed, the dispersion relation in the presence of a uniform background magnetic field takes the form \cite{Neves:2021tbt} %can be obtained directly from ..., 
\begin{eqnarray}
w_{1}\left(\mathbf{k}\right)&=&ck\left[1-\frac{8\alpha^{2}\hbar^{3}}{45m_{e}^{4}c^{5}}\left(\mathbf{\hat{k}}\times\mathbf{B}\right)^{2}\right], \\
w_{2}\left(\mathbf{k}\right)&=&ck\left[1-\frac{14\alpha^{2}\hbar^{3}}{45m_{e}^{4}c^{5}}\left(\mathbf{\hat{k}}\times\mathbf{B}\right)^{2}\right]. 
\end{eqnarray}

Here, the factor $\Delta\Lambda$ is given by
\begin{eqnarray}
\Delta\Lambda\approx2+\frac{22\alpha^{2}\hbar^{3}}{15m_{e}^{4}c^{5}}B^{2}sin^{2}\theta,   
\end{eqnarray}
while the effective sigma obtained is
\begin{eqnarray}
\sigma_{eff}=\left(1+\frac{11\alpha^{2}\hbar^{3}}{30m_{e}^{4}c^{5}}B^{2}\right)\sigma.
\end{eqnarray}

%************************************************

\subsection{Generalized Born-Infeld electrodynamics}\label{BI}

The main motivation of Born and Infeld to propose their theory was to ensure the finiteness of the electric field self-energy \cite{Born:1934gh}. Recently, there has been a renewed interest in the Born-Infeld theory in the context of string theory, quantum gravity models, and theories with magnetic monopoles \cite{Fradkin:1985qd,Bergshoeff:1987at,Ayon-Beato:2000mjt,Gibbons:2001gy,Banerjee:2012zm,Gunasekaran:2012dq,NiauAkmansoy:2017kbw,NiauAkmansoy:2018ilv}. 

The generalized Born-Infeld Lagrangian density is given by\cite{Kruglov:2009he,Gaete:2014nda}
\begin{eqnarray}\label{LBI}
\mathcal{L}_{BI}\left(\mathcal{F},\mathcal{G}\right)=b^{2}\left[1-\left(1-2\frac{\mathcal{F}}{b^{2}}-\frac{\mathcal{G}^{2}}{b^{4}}\right)^{p}\right], 
\end{eqnarray}
where $b$ is a scale parameter and $p$ is a real number $0<{p}<1$. The standard Born-Infeld electrodynamics is recovered when $p=1/2$.

Following the procedure described in Sec. \ref{MDR}, the dispersion relation takes the form\cite{Neves:2021tbt}
\begin{eqnarray}
w_{1}\left({k}\right)&=&ck\sqrt{1-2\left(1-p\right)\frac{\left(\mathbf{\hat{k}\times\mathbf{B}}\right)^{2}}{\mathbf{B}^{2}+b^{2}}}, \\
w_{2}\left({k}\right)&=&ck\sqrt{1-\frac{\left(\mathbf{\hat{k}\times\mathbf{B}}\right)^{2}}{\mathbf{B}^{2}+b^{2}}}. 
\end{eqnarray}

Now, considering the particular case $p=1/2$, both frequencies $w_{1}$ and $w_{2}$ reduce, in the weak field limit, i.e., $b\gg{B}$, to
\begin{eqnarray}\label{frequencyBIweak}
w\left({k}\right)&=&ck\sqrt{1-\frac{B^{2}}{b^{2}}sin^{2}\theta}.
\end{eqnarray}

The frequencies above are always real. It is important to note that in order to derive the spectral distribution, the frequencies are constrained to be real and positive definite. 

In the framework of Born-Infeld theory, the factor $\Delta\Lambda$ assumes the form
\begin{eqnarray}\label{deltaBI}
\Delta\Lambda\approx2+3\frac{B^{2}}{b^{2}}sin^{2}\theta.    
\end{eqnarray}

The effective sigma, in turn, is given by
\begin{eqnarray}\label{sigmaBI}
%\sigma_{eff}=\left(1+\frac{B^{2}}{b^{2}}\right)\sigma.   
%
\sigma_{eff}=\left(1+\frac{3B^{2}}{4b^{2}}\right)\sigma.   \end{eqnarray}

As commented above, from the nonlinear electrodynamics perspective, the relations above indicate to us that the quantities related to the blackbody radiation and to the thermodynamics parameters depend on the magnitude of the uniform external magnetic field, the angle between the wave vector and the background magnetic field, and the parameters of each specific nonlinear model. 

%*************************************************************

\subsection{Logarithmic electrodynamics}\label{Log}

Another nonlinear model we intend to explore is the logarithmic electrodynamics \cite{Gaete:2013dta}, where the Lagrangian density is given by
\begin{eqnarray}
\mathcal{L}_{ln}\left(\mathcal{F},\mathcal{G}\right)=-\beta^{2}ln\left[1-\frac{\mathcal{F}}{\beta^{2}}-\frac{\mathcal{G}^{2}}{2\beta^{4}}\right].
\end{eqnarray}

Maxwell electromagnetism is recovered in the limit $\beta\rightarrow\infty$. The dispersion relation for each mode yields \cite{Neves:2021tbt}
\begin{eqnarray}
w_{1}\left({\mathbf{k}}\right)&=&ck\sqrt{1-\frac{2\left(\mathbf{\hat{k}}\times\mathbf{B}\right)^{2}}{2\beta^{2}+\mathbf{B}^{2}}}, \\
w_{2}\left(\mathbf{k}\right)&=&ck\sqrt{1-\frac{\left(\mathbf{\hat{k}\times\mathbf{B}}\right)^{2}}{\mathbf{B}^{2}+\beta^{2}}}. 
\end{eqnarray}

To ensure that the energy density is positive definite, the condition $B<\sqrt{2}\beta$ must be satisfied\cite{Neves:2021tbt}.

In the weak field limit ($\beta\gg{B}$), both frequencies reduce to (\ref{frequencyBIweak}), which provides the same frequency modes as the Born-Infeld theory. 
Therefore, the logarithmic electrodynamics does not show birefringence in the mentioned regime. In addition, $\Delta\Lambda$ and $\sigma_{eff}$ are given by relations (\ref{deltaBI}) and (\ref{sigmaBI}), respectively.

%************************************************
\subsection{Electromagnetic wave propagation perpendicular to the background magnetic field}\label{perpendicularpropagation}

In what follows, we will take into account the special case where the electromagnetic waves propagate perpendicular to a uniform external magnetic field. In this configuration, we have $\theta=\pi/2$, and the coefficients $\Lambda_{1,2}$ in relations (\ref{lambda1}) and (\ref{lambda2}) reduce to
\begin{eqnarray}
\Lambda_{1}&=&\sqrt{1-\frac{d_{1}}{c_{1}}B^{2}}, \\
\Lambda_{2}&=&\sqrt{1-\frac{d_{2}B^{2}}{c_{1}+d_{2}{B}^{2}}}. 
\end{eqnarray}

As a consequence, the factor $\Delta\Lambda$ depends only on the magnitude of the background magnetic field $B$ and assumes the form 
\begin{eqnarray}
\Delta\Lambda\left(B\right)=\left(1-\frac{d_{1}}{c_{1}}B^{2}\right)^{-3/2}+\left(1-\frac{d_{2}B^{2}}{c_{1}+d_{2}{B}^{2}}\right)^{-3/2}. 
\end{eqnarray}

Now, with the mentioned assumptions, the task of solving the integral (\ref{logZ}) is easier since there is no angular dependence anymore. In this context, it allows us to probe the regime of strong magnetic fields, i.e., one only needs to ensure that the conditions $c_{1}>{d_{1}}B^{2}$ and $1-{d_{2}B^{2}}/\left(c_{1}+d_{2}B^{2}\right)>0$
are satisfied. In such a case, the spectral energy density distribution reads as
\begin{eqnarray}\label{energydensity}
u\left(\nu,T\right)=\left(\frac{4\pi\nu^{2}\Delta\Lambda\left(B\right)}{c^{3}}\right)\frac{h\nu}{\left(e^{\beta{h}\nu}-1\right)}.
\end{eqnarray}

With regard to the effective sigma, since there is no angular dependence anymore, i.e., the anisotropy in the energy density distribution vanishes, one promptly obtains
\begin{eqnarray}\label{effectivesigmakperpendicularB}
%\sigma_{eff}=\left(\frac{\pi^{2}k_{B}^{4}}{60\hbar^{3}c^{2}}\right)\frac{\Delta\Lambda\left(B\right)}{2}.
%
\sigma_{eff}=\left(\frac{\sigma}{2}\right)\Delta\Lambda\left(B\right).
\end{eqnarray}

Note that in the absence of angular dependence, the energy density and the radiance are related to each other by purely geometric factors.

In lab setups, it is feasible to arrange the physical system in such a way that one has an electromagnetic wave propagating perpendicular to a uniform external magnetic field. On the other hand, with respect to astronomical observations, where the luminosity coming from compact objects can be measured, for instance, this condition is a very restrictive one. 

We also call attention to the fact that the in ultrastrong field regime, i.e., whenever $B\gg{b}$ in Born-Infeld theory, or $B\gg\beta$ in logarithmic electrodynamics, the frequencies do not depend on the magnitude of the external magnetic field, relying only on the angular variable $\theta$ \cite{Neves:2021tbt}. In this regime, our framework cannot be applied.  

In the next section, the factor $\Delta\Lambda$ and the effective sigma $\sigma_{eff}$ will be computed in the regime of strong fields assuming magnetic field intensities of the order of the critical magnetic field $\varepsilon_{c}$.

%*******************************
\subsubsection{Born-Infeld theory}

Let us then compute the above quantities for the Born-Infeld theory under the conditions above mentioned. Taking into account the Lagrangian (\ref{LBI}), both the frequencies for $p=1/2$ read as
\begin{eqnarray}
w\left({k}\right)&=&ck\sqrt{1-\frac{B^{2}}{{B}^{2}+b^{2}}}, 
\end{eqnarray}
which gives us
\begin{eqnarray}
\Delta\Lambda=2\frac{\left(B^{2}+b^{2}\right)^{3/2}}{b^{3}}.    
\end{eqnarray}

Concerning the factor $\Delta\Lambda$, if one assumes, in units of $\hbar=c=k_{B}=1$, a magnetic field intensity $B=3MeV^{2}$ and $b=3MeV^{2}$, one obtains $\Delta\Lambda\approx5,65$. In this scenario, the effective Stefan-Boltzmann constant takes the value $\sigma_{eff}=2,82\sigma$. The number of accessible states, on the other hand, allows $\Delta\Lambda/2\approx2,82$ more photons to each frequency mode.

%************************************************

\subsubsection{Logarithmic electrodynamics}

The logarithmic electrodynamics is induced by radiative corrections in the regime of slowly varying fields, which increases logarithmically with the field strengths even in the regime of strong field intensities. As a consequence, the logarithmic Lagrangian is valid for values of electric and magnetic fields stronger than the critical value.

Performing the computation of the frequencies for each mode, we obtain\cite{Neves:2021tbt}
\begin{eqnarray}
w_{1}\left({{k}}\right)&=&ck\sqrt{1-\frac{2B^{2}}{2\beta^{2}+{B}^{2}}}, \\
w_{2}\left({k}\right)&=&ck\sqrt{1-\frac{B^{2}}{{B}^{2}+\beta^{2}}}. 
\end{eqnarray} 

From the above frequencies, it should be clear that the Born-Infeld and the logarithmic electrodynamics differ in the regime of strong fields. In logarithmic electrodynamics, there is the appearance of birefringence, a phenomenon that is absent in the Born-Infeld theory. 

Next, following the same procedure as before, the factor $\Delta\Lambda$ takes the form
\begin{eqnarray}
\Delta\Lambda=\left(\frac{2\beta^{2}-B^{2}}{2\beta^{2}+B^{2}}\right)^{-3/2}+\left(\frac{\beta^{2}}{\beta^{2}+B^{2}}\right)^{-3/2}. \end{eqnarray}

Let us then assume a magnetic field $B=3MeV^{2}$ and $\beta=3MeV^{2}$. These values provide us with a factor $\Delta\Lambda\approx8$. The effective Stefan-Boltzmann constant, in turn, takes the value $\sigma_{eff}\approx4\sigma$ and $N_{LE}\approx4N$.

We would like to highlight that for magnetic fields $B$ with an intensity greater than $\sqrt{2}\beta$, there will be the emergence of imaginary terms in the frequency modes. In this respect, the electromagnetic waves will be attenuated and will not contribute to the thermalization process, leading to no contribution to the emission frequency spectrum.   

Before concluding this section, we also would like to point out that the computations of the frequencies in the Euler-Heisenberg electrodynamics in the strong magnetic field regime are more complicated since they involve the full nonperturbative effective action (\ref{LEH}), and it will not be considered here.
 
%************************************************
\subsection{Some further remarks about blackbody radiation in nonlinear electrodynamics}\label{someremarks}

Let us now discuss some further consequences of the nonlinearity in the thermodynamics of blackbody radiation. From the preceding sections, we have shown that the parameter that carries information about the nonlinearity of the magnetic field, %value of 
$\Delta\Lambda$, is always greater than $2$ in the analyzed models, leading to a modification in the value of the Stefan-Boltzmann constant (\ref{effectivesigma}). As a consequence, the energy density of the photon gas (\ref{totalenergydensity}), for instance, will have more stored energy than in Maxwell electrodynamics. Physically, the photon propagation in the background magnetic field leads to an energy transfer to the photon gas, increasing, in this way, its energy. Analogously, the pressure, entropy, and heat capacity densities associated with the photon ideal gas in Eqs. (\ref{pressure}) and (\ref{heat}) will increase as well. 

\begin{center}
\begin{figure}[htb]
\begin{minipage}{0.5\textwidth}
\begin{tikzpicture}
  \node (img)  {\includegraphics[scale=0.65]{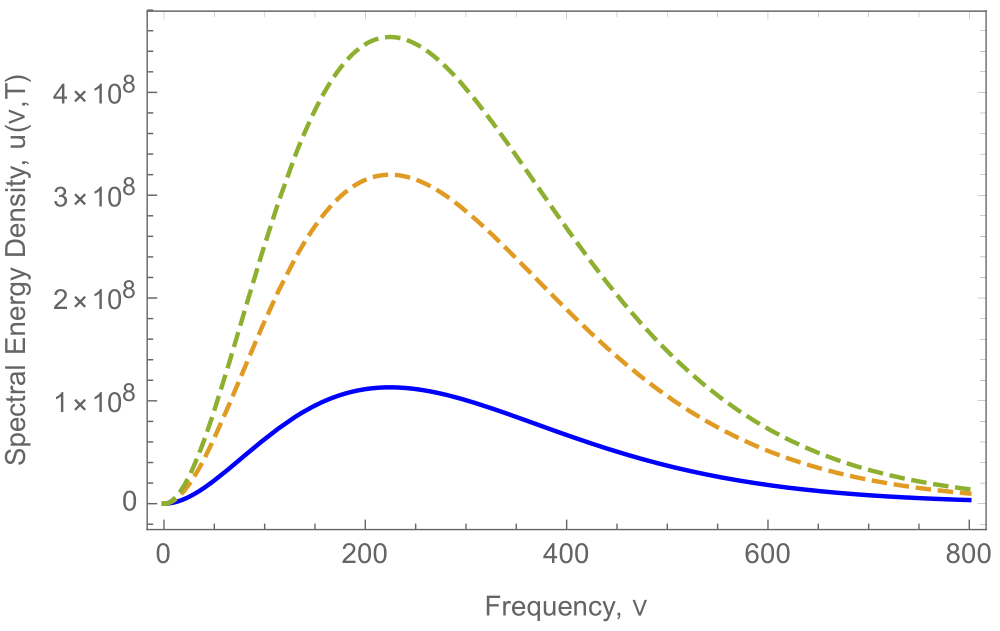}};
  %\node[below=of img, node distance=0cm, yshift=1cm,font={\color{black}}] {Frequency, $\nu$ ($\times10^{2}$) };
  %\node[left=of img, node distance=0cm, rotate=90, anchor=center,yshift=-0.7cm,font={\color{black}}] {Spectral Density, u ($\times10^{8}$)};
 \end{tikzpicture}
\end{minipage}%
\caption{Graph of the spectral energy density distribution of the evaluated models for $T=0.5keV$ when the photon propagation is perpendicular to the external magnetic field. Here, we adopted $c=\hbar=k_{B}=1$. The conversion of Tesla $T$ to the natural system is $1T=6.8\times10^{-16}GeV^{2}$. In addition, in each model, we have considered a background magnetic field intensity $B=3MeV^{2}$. For both Born-Infeld and logarithmic electrodynamics, we set $\beta=3MeV^{2}$ and $b=3MeV^{2}$.  The blue line corresponds to the Planck spectrum, while the dashed orange and green are associated with Born-Infeld and logarithmic electrodynamics, respectively. According to Wien's law, $\nu_{max}\approx0.45T$, and the peak is localized at $\nu_{max}\approx225eV$.}\label{figure}
\end{figure}
\end{center}

With regard to the spectral density deviations due to nonlinearity, we plotted, in Fig. \ref{figure}, the frequency spectrum arising from the Maxwell theory and from the Born-Infeld and logarithmic electrodynamics when the photon propagation is perpendicular to the background magnetic field. The graph shows that, for a given temperature, the nonlinear models present an increase in the blackbody curve in comparison to the standard Planck distribution. This fact can be understood by evaluating the density of states $g\left(\nu,B\right)$, which is given by
\begin{eqnarray}\label{densitystates}
g\left(\nu,B\right)=\frac{4\pi{\nu^{2}}}{c^{3}}\Delta\Lambda\left(B\right).
\end{eqnarray}
%Indeed, 

According to relation (\ref{densitystates}), the nonlinearities of the magnetic field lead to more accessible states to the photon gas and cause an increase in the number of photons for each frequency. 

To conclude this section, we remark that all the models under consideration reduce to a similar form in the weak field approximation, i.e., all models have corrections to the usual Maxwell term $\mathcal{F}$ which are proportional to $\mathcal{F}^{2}$ and $\mathcal{G}^{2}$. It guarantees that the constraints $c_{1}\gg{d_{1}}\left(\mathbf{\hat{k}}\times\mathbf{B}\right)^{2}$ and $c_{1}\gg-{d_{2}}\left(\mathbf{\hat{k}}\cdot\mathbf{B}\right)^{2}$ are satisfied in order to take into account the angular dependence. 

%*************************************

\section{Final Remarks}\label{conclusions}

In this paper, we have investigated the consequences of electromagnetic waves propagating in a magnetized medium. Specifically, we have derived the blackbody radiation laws in this situation, such as the spectral distribution and the Stefan-Boltzmann law. Particularly, we have included the angular dependence at the frequency spectrum, which has shown us the emergence of a quadrupole term in the regime of weak fields. The Rayleigh-Jeans formula was contemplated as well. Furthermore, we have shown that the connection between the energy density and the radiance by geometric factors is lost in the weak field approximation, while it is restored at the strong one. The modified free energy led to small deviations in the thermodynamic quantities. We also studied the free energy, as well as the energy, pressure, entropy, and heat capacity densities. As an application, we have considered three distinct nonlinear electrodynamics, namely, the Euler-Heisenberg, the generalized Born-Infeld, and the logarithmic electrodynamics. The strong field regime was also probed in the particular case when the wave propagation is perpendicular to the background magnetic field. We would like to remark that our approach can be used with any nonlinear electrodynamics model within the validity of our assumptions. On the other hand, our framework does not treat the self-interaction of the photons rigorously, only effectively. A way to generalize this framework and take the photon self-interaction into account could be by considering the procedures of field theory at finite temperature \cite{Braaten:1995cm}.

As a future prospect, we intend to extend our analysis and investigate the thermodynamics of blackbody radiation in Lorentz symmetry violating theories in connection with nonlinear electrodynamics. Such scenarios seem plausible in neutron stars with strong magnetic fields, which could, in principle, unveil phenomena of physics beyond the SM. Features related to the blackbody phenomenon in compact extra dimensions similar to what has been done by Ramos\cite{Ramos:2014mda} can also be contemplated. In this sense, it might be worthwhile to explore nonlinear models which depend exclusively on powers of $\mathcal{F}$ and then study the role of the extra dimensions in the blackbody radiation.

Finally, we would like to stress that there is intense research in modeling the emission spectrum of magnetars in the region of soft x rays. Magnetars are neutron stars having extreme magnetic field intensity of the order of $10^{11}T$. The study of their spectrum can be valuable to understanding features related to the powerful magnetic field in such compact objects. Usually, the emission spectrum is modeled by taking into account a superposition of two blackbody components or a blackbody plus a power law model. A computational implementation of our results and the use of the observational data from the Chandra X-ray Observatory, XMM-Newton, and Suzaku, can be very promising and have the potential to improve the observed x-ray luminosity of magnetars as well as be used to test the linearity of Maxwell theory and to set constraints on nonlinear electrodynamic models.
A manner to incorporate magnetars in our formalism and study features related to the magnetic fields of such objects could be by taking into account the Born-Infeld theory, for instance. In this sense, recent observations of the hydrogen atom spectrum from the Born-Infeld electrodynamics perspective suggest that the scale factor $b$ should be larger than $10^{11}T$\cite{Franklin:2011zz,NiauAkmansoy:2017kbw}, while measurements of light by light scattering at the LHC in Pb-Pb collisions would restrict the scale factor $b$ to be larger than $10^{19}T$\cite{Ellis:2017edi}. Nevertheless, because of the kinematic cuts in ATLAS analysis, smaller values of the scale factor $b$ cannot be reached\cite{Ellis:2017edi}, making it possible, in this way, to use magnetars to probe the range $10^{11}T\le{b}\le10^{19}T$ in both weak and strong regimes of our formalism.

Last but not least, we would like to call attention to the fact that the nonlinearity in the regime of strong magnetic fields can have an important role in the physical properties of magnetars during the cooling process, impacting the internal structure of these objects, such as the equation of state of the dense matter, superfluidity of several baryon species and the neutrino emission mechanisms. In this sense, a distinct luminosity pattern would be expected coming from magnetars in comparison to neutron stars, which could be useful to distinguish such objects, besides providing valuable information about the interior of magnetars. We hope that these interesting features will stimulate further work on the subject.

%****************************************
\section{Acknowledgments}

This work is a part of the project INCT-FNA proc. No. 464898/2014-5. R.T. acknowledges financial support from the PCI program of the Brazilian agency Conselho Nacional de Desenvolvimento Científico e Tecnológico – CNPq. S.B.D. thanks CNPq for partial financial support. %The authors also acknowledge the anonymous referee for pointing out features related to the conformal equivalence between the Boillat-Birula and Novello metrics.

\bibliographystyle{apsrev4-1}
\bibliography{Bibliography.bib}

%merlin.mbs apsrev4-1.bst 2010-07-25 4.21a (PWD, AO, DPC) hacked
%Control: key (0)
%Control: author (72) initials jnrlst
%Control: editor formatted (1) identically to author
%Control: production of article title (-1) disabled
%Control: page (0) single
%Control: year (1) truncated
%Control: production of eprint (0) enabled
\begin{thebibliography}{68}%
\makeatletter
\providecommand \@ifxundefined [1]{%
 \@ifx{#1\undefined}
}%
\providecommand \@ifnum [1]{%
 \ifnum #1\expandafter \@firstoftwo
 \else \expandafter \@secondoftwo
 \fi
}%
\providecommand \@ifx [1]{%
 \ifx #1\expandafter \@firstoftwo
 \else \expandafter \@secondoftwo
 \fi
}%
\providecommand \natexlab [1]{#1}%
\providecommand \enquote  [1]{``#1''}%
\providecommand \bibnamefont  [1]{#1}%
\providecommand \bibfnamefont [1]{#1}%
\providecommand \citenamefont [1]{#1}%
\providecommand \href@noop [0]{\@secondoftwo}%
\providecommand \href [0]{\begingroup \@sanitize@url \@href}%
\providecommand \@href[1]{\@@startlink{#1}\@@href}%
\providecommand \@@href[1]{\endgroup#1\@@endlink}%
\providecommand \@sanitize@url [0]{\catcode `\\12\catcode `\$12\catcode
  `\&12\catcode `\#12\catcode `\^12\catcode `\_12\catcode `\%12\relax}%
\providecommand \@@startlink[1]{}%
\providecommand \@@endlink[0]{}%
\providecommand \url  [0]{\begingroup\@sanitize@url \@url }%
\providecommand \@url [1]{\endgroup\@href {#1}{\urlprefix }}%
\providecommand \urlprefix  [0]{URL }%
\providecommand \Eprint [0]{\href }%
\providecommand \doibase [0]{http://dx.doi.org/}%
\providecommand \selectlanguage [0]{\@gobble}%
\providecommand \bibinfo  [0]{\@secondoftwo}%
\providecommand \bibfield  [0]{\@secondoftwo}%
\providecommand \translation [1]{[#1]}%
\providecommand \BibitemOpen [0]{}%
\providecommand \bibitemStop [0]{}%
\providecommand \bibitemNoStop [0]{.\EOS\space}%
\providecommand \EOS [0]{\spacefactor3000\relax}%
\providecommand \BibitemShut  [1]{\csname bibitem#1\endcsname}%
\let\auto@bib@innerbib\@empty
%</preamble>
\bibitem [{\citenamefont {Karbstein}(2020)}]{Karbstein:2019oej}%
  \BibitemOpen
  \bibfield  {author} {\bibinfo {author} {\bibfnamefont {F.}~\bibnamefont
  {Karbstein}},\ }\href {\doibase 10.3390/particles3010005} {\bibfield
  {journal} {\bibinfo  {journal} {Particles}\ }\textbf {\bibinfo {volume}
  {3}},\ \bibinfo {pages} {39} (\bibinfo {year} {2020})},\ \Eprint
  {http://arxiv.org/abs/1912.11698} {arXiv:1912.11698 [hep-ph]} \BibitemShut
  {NoStop}%
\bibitem [{\citenamefont {Marklund}\ and\ \citenamefont
  {Lundin}(2009)}]{Marklund:2008gj}%
  \BibitemOpen
  \bibfield  {author} {\bibinfo {author} {\bibfnamefont {M.}~\bibnamefont
  {Marklund}}\ and\ \bibinfo {author} {\bibfnamefont {J.}~\bibnamefont
  {Lundin}},\ }\href {\doibase 10.1140/epjd/e2009-00169-6} {\bibfield
  {journal} {\bibinfo  {journal} {Eur. Phys. J. D}\ }\textbf {\bibinfo {volume}
  {55}},\ \bibinfo {pages} {319} (\bibinfo {year} {2009})},\ \Eprint
  {http://arxiv.org/abs/0812.3087} {arXiv:0812.3087 [hep-th]} \BibitemShut
  {NoStop}%
\bibitem [{\citenamefont {Battesti}\ and\ \citenamefont
  {Rizzo}(2013)}]{Battesti:2012hf}%
  \BibitemOpen
  \bibfield  {author} {\bibinfo {author} {\bibfnamefont {R.}~\bibnamefont
  {Battesti}}\ and\ \bibinfo {author} {\bibfnamefont {C.}~\bibnamefont
  {Rizzo}},\ }\href {\doibase 10.1088/0034-4885/76/1/016401} {\bibfield
  {journal} {\bibinfo  {journal} {Rept. Prog. Phys.}\ }\textbf {\bibinfo
  {volume} {76}},\ \bibinfo {pages} {016401} (\bibinfo {year} {2013})},\
  \Eprint {http://arxiv.org/abs/1211.1933} {arXiv:1211.1933 [physics.optics]}
  \BibitemShut {NoStop}%
\bibitem [{\citenamefont {King}\ and\ \citenamefont
  {Heinzl}(2016)}]{King:2015tba}%
  \BibitemOpen
  \bibfield  {author} {\bibinfo {author} {\bibfnamefont {B.}~\bibnamefont
  {King}}\ and\ \bibinfo {author} {\bibfnamefont {T.}~\bibnamefont {Heinzl}},\
  }\href {\doibase 10.1017/hpl.2016.1} {\bibfield  {journal} {\bibinfo
  {journal} {High Power Laser Sci. Eng.}\ }\textbf {\bibinfo {volume} {4}}
  (\bibinfo {year} {2016}),\ 10.1017/hpl.2016.1},\ \Eprint
  {http://arxiv.org/abs/1510.08456} {arXiv:1510.08456 [hep-ph]} \BibitemShut
  {NoStop}%
\bibitem [{\citenamefont {Guzman-Herrera}\ and\ \citenamefont
  {Breton}(2021)}]{Guzman-Herrera:2020ffc}%
  \BibitemOpen
  \bibfield  {author} {\bibinfo {author} {\bibfnamefont {E.}~\bibnamefont
  {Guzman-Herrera}}\ and\ \bibinfo {author} {\bibfnamefont {N.}~\bibnamefont
  {Breton}},\ }\href {\doibase 10.1140/epjc/s10052-020-08783-1} {\bibfield
  {journal} {\bibinfo  {journal} {Eur. Phys. J. C}\ }\textbf {\bibinfo {volume}
  {81}},\ \bibinfo {pages} {115} (\bibinfo {year} {2021})},\ \Eprint
  {http://arxiv.org/abs/2008.10739} {arXiv:2008.10739 [hep-th]} \BibitemShut
  {NoStop}%
\bibitem [{\citenamefont {Kim}(2022)}]{Kim:2022xum}%
  \BibitemOpen
  \bibfield  {author} {\bibinfo {author} {\bibfnamefont {J.~Y.}\ \bibnamefont
  {Kim}},\ }\href {\doibase 10.1140/epjc/s10052-022-10435-5} {\bibfield
  {journal} {\bibinfo  {journal} {Eur. Phys. J. C}\ }\textbf {\bibinfo {volume}
  {82}},\ \bibinfo {pages} {485} (\bibinfo {year} {2022})},\ \Eprint
  {http://arxiv.org/abs/2202.11913} {arXiv:2202.11913 [gr-qc]} \BibitemShut
  {NoStop}%
\bibitem [{\citenamefont {Mosquera~Cuesta}\ and\ \citenamefont
  {Salim}(2004)}]{MosqueraCuesta:2004em}%
  \BibitemOpen
  \bibfield  {author} {\bibinfo {author} {\bibfnamefont {H.~J.}\ \bibnamefont
  {Mosquera~Cuesta}}\ and\ \bibinfo {author} {\bibfnamefont {J.~M.}\
  \bibnamefont {Salim}},\ }\href {\doibase 10.1111/j.1365-2966.2004.08375.x}
  {\bibfield  {journal} {\bibinfo  {journal} {Mon. Not. Roy. Astron. Soc.}\
  }\textbf {\bibinfo {volume} {354}},\ \bibinfo {pages} {L55} (\bibinfo {year}
  {2004})},\ \Eprint {http://arxiv.org/abs/astro-ph/0403045}
  {arXiv:astro-ph/0403045} \BibitemShut {NoStop}%
\bibitem [{\citenamefont {Baggioli}\ and\ \citenamefont
  {Pujolas}(2016)}]{Baggioli:2016oju}%
  \BibitemOpen
  \bibfield  {author} {\bibinfo {author} {\bibfnamefont {M.}~\bibnamefont
  {Baggioli}}\ and\ \bibinfo {author} {\bibfnamefont {O.}~\bibnamefont
  {Pujolas}},\ }\href {\doibase 10.1007/JHEP12(2016)107} {\bibfield  {journal}
  {\bibinfo  {journal} {JHEP}\ }\textbf {\bibinfo {volume} {12}},\ \bibinfo
  {pages} {107} (\bibinfo {year} {2016})},\ \Eprint
  {http://arxiv.org/abs/1604.08915} {arXiv:1604.08915 [hep-th]} \BibitemShut
  {NoStop}%
\bibitem [{\citenamefont {Bi}\ and\ \citenamefont {Tao}(2021)}]{Bi:2021maw}%
  \BibitemOpen
  \bibfield  {author} {\bibinfo {author} {\bibfnamefont {S.}~\bibnamefont
  {Bi}}\ and\ \bibinfo {author} {\bibfnamefont {J.}~\bibnamefont {Tao}},\
  }\href {\doibase 10.1007/JHEP06(2021)174} {\bibfield  {journal} {\bibinfo
  {journal} {JHEP}\ }\textbf {\bibinfo {volume} {06}},\ \bibinfo {pages} {174}
  (\bibinfo {year} {2021})},\ \Eprint {http://arxiv.org/abs/2101.00912}
  {arXiv:2101.00912 [hep-th]} \BibitemShut {NoStop}%
\bibitem [{\citenamefont {Sorokin}(2022)}]{Sorokin:2021tge}%
  \BibitemOpen
  \bibfield  {author} {\bibinfo {author} {\bibfnamefont {D.~P.}\ \bibnamefont
  {Sorokin}},\ }\href {\doibase 10.1002/prop.202200092} {\bibfield  {journal}
  {\bibinfo  {journal} {Fortsch. Phys.}\ }\textbf {\bibinfo {volume} {70}},\
  \bibinfo {pages} {2200092} (\bibinfo {year} {2022})},\ \Eprint
  {http://arxiv.org/abs/2112.12118} {arXiv:2112.12118 [hep-th]} \BibitemShut
  {NoStop}%
\bibitem [{\citenamefont {Workman}\ \emph {et~al.}(2022)\citenamefont {Workman}
  \emph {et~al.}}]{ParticleDataGroup:2022pth}%
  \BibitemOpen
  \bibfield  {author} {\bibinfo {author} {\bibfnamefont {R.~L.}\ \bibnamefont
  {Workman}} \emph {et~al.} (\bibinfo {collaboration} {Particle Data Group}),\
  }\href {\doibase 10.1093/ptep/ptac097} {\bibfield  {journal} {\bibinfo
  {journal} {PTEP}\ }\textbf {\bibinfo {volume} {2022}},\ \bibinfo {pages}
  {083C01} (\bibinfo {year} {2022})}\BibitemShut {NoStop}%
\bibitem [{\citenamefont {Schwinger}(1951)}]{Schwinger:1951nm}%
  \BibitemOpen
  \bibfield  {author} {\bibinfo {author} {\bibfnamefont {J.~S.}\ \bibnamefont
  {Schwinger}},\ }\href {\doibase 10.1103/PhysRev.82.664} {\bibfield  {journal}
  {\bibinfo  {journal} {Phys. Rev.}\ }\textbf {\bibinfo {volume} {82}},\
  \bibinfo {pages} {664} (\bibinfo {year} {1951})}\BibitemShut {NoStop}%
\bibitem [{\citenamefont {Boillat}(1970)}]{Boillat:1970gw}%
  \BibitemOpen
  \bibfield  {author} {\bibinfo {author} {\bibfnamefont {G.}~\bibnamefont
  {Boillat}},\ }\href {\doibase 10.1063/1.1665231} {\bibfield  {journal}
  {\bibinfo  {journal} {J. Math. Phys.}\ }\textbf {\bibinfo {volume} {11}},\
  \bibinfo {pages} {941} (\bibinfo {year} {1970})}\BibitemShut {NoStop}%
\bibitem [{\citenamefont {Bialynicka-Birula}\ and\ \citenamefont
  {Bialynicki-Birula}(1970)}]{Bialynicka-Birula:1970nlh}%
  \BibitemOpen
  \bibfield  {author} {\bibinfo {author} {\bibfnamefont {Z.}~\bibnamefont
  {Bialynicka-Birula}}\ and\ \bibinfo {author} {\bibfnamefont {I.}~\bibnamefont
  {Bialynicki-Birula}},\ }\href {\doibase 10.1103/PhysRevD.2.2341} {\bibfield
  {journal} {\bibinfo  {journal} {Phys. Rev. D}\ }\textbf {\bibinfo {volume}
  {2}},\ \bibinfo {pages} {2341} (\bibinfo {year} {1970})}\BibitemShut
  {NoStop}%
\bibitem [{\citenamefont {Soleng}(1995)}]{Soleng:1995kn}%
  \BibitemOpen
  \bibfield  {author} {\bibinfo {author} {\bibfnamefont {H.~H.}\ \bibnamefont
  {Soleng}},\ }\href {\doibase 10.1103/PhysRevD.52.6178} {\bibfield  {journal}
  {\bibinfo  {journal} {Phys. Rev. D}\ }\textbf {\bibinfo {volume} {52}},\
  \bibinfo {pages} {6178} (\bibinfo {year} {1995})},\ \Eprint
  {http://arxiv.org/abs/hep-th/9509033} {arXiv:hep-th/9509033} \BibitemShut
  {NoStop}%
\bibitem [{\citenamefont {Bronnikov}(2001)}]{Bronnikov:2000vy}%
  \BibitemOpen
  \bibfield  {author} {\bibinfo {author} {\bibfnamefont {K.~A.}\ \bibnamefont
  {Bronnikov}},\ }\href {\doibase 10.1103/PhysRevD.63.044005} {\bibfield
  {journal} {\bibinfo  {journal} {Phys. Rev. D}\ }\textbf {\bibinfo {volume}
  {63}},\ \bibinfo {pages} {044005} (\bibinfo {year} {2001})},\ \Eprint
  {http://arxiv.org/abs/gr-qc/0006014} {arXiv:gr-qc/0006014} \BibitemShut
  {NoStop}%
\bibitem [{\citenamefont {Ayon-Beato}\ and\ \citenamefont
  {Garcia}(2000)}]{Ayon-Beato:2000mjt}%
  \BibitemOpen
  \bibfield  {author} {\bibinfo {author} {\bibfnamefont {E.}~\bibnamefont
  {Ayon-Beato}}\ and\ \bibinfo {author} {\bibfnamefont {A.}~\bibnamefont
  {Garcia}},\ }\href {\doibase 10.1016/S0370-2693(00)01125-4} {\bibfield
  {journal} {\bibinfo  {journal} {Phys. Lett. B}\ }\textbf {\bibinfo {volume}
  {493}},\ \bibinfo {pages} {149} (\bibinfo {year} {2000})},\ \Eprint
  {http://arxiv.org/abs/gr-qc/0009077} {arXiv:gr-qc/0009077} \BibitemShut
  {NoStop}%
\bibitem [{\citenamefont {Cataldo}\ \emph {et~al.}(2000)\citenamefont
  {Cataldo}, \citenamefont {Cruz}, \citenamefont {del Campo},\ and\
  \citenamefont {Garcia}}]{Cataldo:2000we}%
  \BibitemOpen
  \bibfield  {author} {\bibinfo {author} {\bibfnamefont {M.}~\bibnamefont
  {Cataldo}}, \bibinfo {author} {\bibfnamefont {N.}~\bibnamefont {Cruz}},
  \bibinfo {author} {\bibfnamefont {S.}~\bibnamefont {del Campo}}, \ and\
  \bibinfo {author} {\bibfnamefont {A.}~\bibnamefont {Garcia}},\ }\href
  {\doibase 10.1016/S0370-2693(00)00609-2} {\bibfield  {journal} {\bibinfo
  {journal} {Phys. Lett. B}\ }\textbf {\bibinfo {volume} {484}},\ \bibinfo
  {pages} {154} (\bibinfo {year} {2000})},\ \Eprint
  {http://arxiv.org/abs/hep-th/0008138} {arXiv:hep-th/0008138} \BibitemShut
  {NoStop}%
\bibitem [{\citenamefont {Burinskii}\ and\ \citenamefont
  {Hildebrandt}(2002)}]{Burinskii:2002pz}%
  \BibitemOpen
  \bibfield  {author} {\bibinfo {author} {\bibfnamefont {A.}~\bibnamefont
  {Burinskii}}\ and\ \bibinfo {author} {\bibfnamefont {S.~R.}\ \bibnamefont
  {Hildebrandt}},\ }\href {\doibase 10.1103/PhysRevD.65.104017} {\bibfield
  {journal} {\bibinfo  {journal} {Phys. Rev. D}\ }\textbf {\bibinfo {volume}
  {65}},\ \bibinfo {pages} {104017} (\bibinfo {year} {2002})},\ \Eprint
  {http://arxiv.org/abs/hep-th/0202066} {arXiv:hep-th/0202066} \BibitemShut
  {NoStop}%
\bibitem [{\citenamefont {Hassaine}\ and\ \citenamefont
  {Martinez}(2007)}]{Hassaine:2007py}%
  \BibitemOpen
  \bibfield  {author} {\bibinfo {author} {\bibfnamefont {M.}~\bibnamefont
  {Hassaine}}\ and\ \bibinfo {author} {\bibfnamefont {C.}~\bibnamefont
  {Martinez}},\ }\href {\doibase 10.1103/PhysRevD.75.027502} {\bibfield
  {journal} {\bibinfo  {journal} {Phys. Rev. D}\ }\textbf {\bibinfo {volume}
  {75}},\ \bibinfo {pages} {027502} (\bibinfo {year} {2007})},\ \Eprint
  {http://arxiv.org/abs/hep-th/0701058} {arXiv:hep-th/0701058} \BibitemShut
  {NoStop}%
\bibitem [{\citenamefont {Pan}\ \emph {et~al.}(2011)\citenamefont {Pan},
  \citenamefont {Jing},\ and\ \citenamefont {Wang}}]{Pan:2011vi}%
  \BibitemOpen
  \bibfield  {author} {\bibinfo {author} {\bibfnamefont {Q.}~\bibnamefont
  {Pan}}, \bibinfo {author} {\bibfnamefont {J.}~\bibnamefont {Jing}}, \ and\
  \bibinfo {author} {\bibfnamefont {B.}~\bibnamefont {Wang}},\ }\href {\doibase
  10.1103/PhysRevD.84.126020} {\bibfield  {journal} {\bibinfo  {journal} {Phys.
  Rev. D}\ }\textbf {\bibinfo {volume} {84}},\ \bibinfo {pages} {126020}
  (\bibinfo {year} {2011})},\ \Eprint {http://arxiv.org/abs/1111.0714}
  {arXiv:1111.0714 [gr-qc]} \BibitemShut {NoStop}%
\bibitem [{\citenamefont {Hendi}(2012)}]{Hendi:2012zz}%
  \BibitemOpen
  \bibfield  {author} {\bibinfo {author} {\bibfnamefont {S.~H.}\ \bibnamefont
  {Hendi}},\ }\href {\doibase 10.1007/JHEP03(2012)065} {\bibfield  {journal}
  {\bibinfo  {journal} {JHEP}\ }\textbf {\bibinfo {volume} {03}},\ \bibinfo
  {pages} {065} (\bibinfo {year} {2012})},\ \Eprint
  {http://arxiv.org/abs/1405.4941} {arXiv:1405.4941 [hep-th]} \BibitemShut
  {NoStop}%
\bibitem [{\citenamefont {Cembranos}\ \emph {et~al.}(2015)\citenamefont
  {Cembranos}, \citenamefont {de~la Cruz-Dombriz},\ and\ \citenamefont
  {Jarillo}}]{Cembranos:2014hwa}%
  \BibitemOpen
  \bibfield  {author} {\bibinfo {author} {\bibfnamefont {J.~A.~R.}\
  \bibnamefont {Cembranos}}, \bibinfo {author} {\bibfnamefont {A.}~\bibnamefont
  {de~la Cruz-Dombriz}}, \ and\ \bibinfo {author} {\bibfnamefont
  {J.}~\bibnamefont {Jarillo}},\ }\href {\doibase
  10.1088/1475-7516/2015/02/042} {\bibfield  {journal} {\bibinfo  {journal}
  {JCAP}\ }\textbf {\bibinfo {volume} {02}},\ \bibinfo {pages} {042} (\bibinfo
  {year} {2015})},\ \Eprint {http://arxiv.org/abs/1407.4383} {arXiv:1407.4383
  [gr-qc]} \BibitemShut {NoStop}%
\bibitem [{\citenamefont {Gaete}\ and\ \citenamefont
  {Helay\"el-Neto}(2014{\natexlab{a}})}]{Gaete:2013dta}%
  \BibitemOpen
  \bibfield  {author} {\bibinfo {author} {\bibfnamefont {P.}~\bibnamefont
  {Gaete}}\ and\ \bibinfo {author} {\bibfnamefont {J.}~\bibnamefont
  {Helay\"el-Neto}},\ }\href {\doibase 10.1140/epjc/s10052-014-2816-4}
  {\bibfield  {journal} {\bibinfo  {journal} {Eur. Phys. J. C}\ }\textbf
  {\bibinfo {volume} {74}},\ \bibinfo {pages} {2816} (\bibinfo {year}
  {2014}{\natexlab{a}})},\ \Eprint {http://arxiv.org/abs/1312.5157}
  {arXiv:1312.5157 [hep-th]} \BibitemShut {NoStop}%
\bibitem [{\citenamefont {Gaete}\ and\ \citenamefont
  {Helay\"el-Neto}(2014{\natexlab{b}})}]{Gaete:2014nda}%
  \BibitemOpen
  \bibfield  {author} {\bibinfo {author} {\bibfnamefont {P.}~\bibnamefont
  {Gaete}}\ and\ \bibinfo {author} {\bibfnamefont {J.}~\bibnamefont
  {Helay\"el-Neto}},\ }\href {\doibase 10.1140/epjc/s10052-014-3182-y}
  {\bibfield  {journal} {\bibinfo  {journal} {Eur. Phys. J. C}\ }\textbf
  {\bibinfo {volume} {74}},\ \bibinfo {pages} {3182} (\bibinfo {year}
  {2014}{\natexlab{b}})},\ \Eprint {http://arxiv.org/abs/1408.3363}
  {arXiv:1408.3363 [hep-th]} \BibitemShut {NoStop}%
\bibitem [{\citenamefont {Gaete}(2016)}]{Gaete:2015qda}%
  \BibitemOpen
  \bibfield  {author} {\bibinfo {author} {\bibfnamefont {P.}~\bibnamefont
  {Gaete}},\ }\href {\doibase 10.1155/2016/2463203} {\bibfield  {journal}
  {\bibinfo  {journal} {Adv. High Energy Phys.}\ }\textbf {\bibinfo {volume}
  {2016}},\ \bibinfo {pages} {2463203} (\bibinfo {year} {2016})},\ \Eprint
  {http://arxiv.org/abs/1509.02594} {arXiv:1509.02594 [hep-th]} \BibitemShut
  {NoStop}%
\bibitem [{\citenamefont {Kruglov}(2014)}]{Kruglov:2014hpa}%
  \BibitemOpen
  \bibfield  {author} {\bibinfo {author} {\bibfnamefont {S.~I.}\ \bibnamefont
  {Kruglov}},\ }\href {\doibase 10.1016/j.aop.2014.12.001} {\bibfield
  {journal} {\bibinfo  {journal} {Annals Phys.}\ }\textbf {\bibinfo {volume}
  {353}},\ \bibinfo {pages} {299} (\bibinfo {year} {2014})},\ \Eprint
  {http://arxiv.org/abs/1410.0351} {arXiv:1410.0351 [physics.gen-ph]}
  \BibitemShut {NoStop}%
\bibitem [{\citenamefont {Kruglov}(2017)}]{Kruglov:2015kda}%
  \BibitemOpen
  \bibfield  {author} {\bibinfo {author} {\bibfnamefont {S.~I.}\ \bibnamefont
  {Kruglov}},\ }\href {\doibase 10.1142/S0218271817500754} {\bibfield
  {journal} {\bibinfo  {journal} {Int. J. Mod. Phys. D}\ }\textbf {\bibinfo
  {volume} {26}},\ \bibinfo {pages} {1750075} (\bibinfo {year} {2017})},\
  \Eprint {http://arxiv.org/abs/1510.06704} {arXiv:1510.06704 [physics.gen-ph]}
  \BibitemShut {NoStop}%
\bibitem [{\citenamefont {Kruglov}(2016)}]{Kruglov:2016ezw}%
  \BibitemOpen
  \bibfield  {author} {\bibinfo {author} {\bibfnamefont {S.~I.}\ \bibnamefont
  {Kruglov}},\ }\href {\doibase 10.1002/andp.201600027} {\bibfield  {journal}
  {\bibinfo  {journal} {Annalen Phys.}\ }\textbf {\bibinfo {volume} {528}},\
  \bibinfo {pages} {588} (\bibinfo {year} {2016})},\ \Eprint
  {http://arxiv.org/abs/1607.07726} {arXiv:1607.07726 [gr-qc]} \BibitemShut
  {NoStop}%
\bibitem [{\citenamefont {Liu}\ \emph {et~al.}(2020)\citenamefont {Liu},
  \citenamefont {Mai}, \citenamefont {Li},\ and\ \citenamefont
  {L\"u}}]{Liu:2019rib}%
  \BibitemOpen
  \bibfield  {author} {\bibinfo {author} {\bibfnamefont {H.-S.}\ \bibnamefont
  {Liu}}, \bibinfo {author} {\bibfnamefont {Z.-F.}\ \bibnamefont {Mai}},
  \bibinfo {author} {\bibfnamefont {Y.-Z.}\ \bibnamefont {Li}}, \ and\ \bibinfo
  {author} {\bibfnamefont {H.}~\bibnamefont {L\"u}},\ }\href {\doibase
  10.1007/s11433-019-1446-1} {\bibfield  {journal} {\bibinfo  {journal} {Sci.
  China Phys. Mech. Astron.}\ }\textbf {\bibinfo {volume} {63}},\ \bibinfo
  {pages} {240411} (\bibinfo {year} {2020})},\ \Eprint
  {http://arxiv.org/abs/1907.10876} {arXiv:1907.10876 [hep-th]} \BibitemShut
  {NoStop}%
\bibitem [{\citenamefont {Gullu}\ and\ \citenamefont
  {Mazharimousavi}(2021)}]{Gullu:2020ant}%
  \BibitemOpen
  \bibfield  {author} {\bibinfo {author} {\bibfnamefont {I.}~\bibnamefont
  {Gullu}}\ and\ \bibinfo {author} {\bibfnamefont {S.~H.}\ \bibnamefont
  {Mazharimousavi}},\ }\href {\doibase 10.1088/1402-4896/abe498} {\bibfield
  {journal} {\bibinfo  {journal} {Phys. Scripta}\ }\textbf {\bibinfo {volume}
  {96}},\ \bibinfo {pages} {045217} (\bibinfo {year} {2021})},\ \Eprint
  {http://arxiv.org/abs/2009.08665} {arXiv:2009.08665 [gr-qc]} \BibitemShut
  {NoStop}%
\bibitem [{\citenamefont {Balakin}\ \emph {et~al.}(2038)\citenamefont
  {Balakin}, \citenamefont {Bochkarev},\ and\ \citenamefont
  {Nizamieva}}]{Balakin:2021arf}%
  \BibitemOpen
  \bibfield  {author} {\bibinfo {author} {\bibfnamefont {A.~B.}\ \bibnamefont
  {Balakin}}, \bibinfo {author} {\bibfnamefont {V.~V.}\ \bibnamefont
  {Bochkarev}}, \ and\ \bibinfo {author} {\bibfnamefont {A.~F.}\ \bibnamefont
  {Nizamieva}},\ }\href {\doibase 10.3390/sym13112038} {\bibfield  {journal}
  {\bibinfo  {journal} {Symmetry}\ }\textbf {\bibinfo {volume} {2021}},\
  \bibinfo {pages} {13} (\bibinfo {year} {2038})},\ \Eprint
  {http://arxiv.org/abs/2110.03005} {arXiv:2110.03005 [gr-qc]} \BibitemShut
  {NoStop}%
\bibitem [{\citenamefont {Balakin}\ and\ \citenamefont
  {Galimova}(2021)}]{Balakin:2021jby}%
  \BibitemOpen
  \bibfield  {author} {\bibinfo {author} {\bibfnamefont {A.~B.}\ \bibnamefont
  {Balakin}}\ and\ \bibinfo {author} {\bibfnamefont {A.~A.}\ \bibnamefont
  {Galimova}},\ }\href {\doibase 10.1103/PhysRevD.104.044059} {\bibfield
  {journal} {\bibinfo  {journal} {Phys. Rev. D}\ }\textbf {\bibinfo {volume}
  {104}},\ \bibinfo {pages} {044059} (\bibinfo {year} {2021})},\ \Eprint
  {http://arxiv.org/abs/2106.01417} {arXiv:2106.01417 [gr-qc]} \BibitemShut
  {NoStop}%
\bibitem [{\citenamefont {Dehghani}\ \emph {et~al.}(2022)\citenamefont
  {Dehghani}, \citenamefont {Setare},\ and\ \citenamefont
  {Zarepour}}]{Dehghani:2021fwb}%
  \BibitemOpen
  \bibfield  {author} {\bibinfo {author} {\bibfnamefont {A.}~\bibnamefont
  {Dehghani}}, \bibinfo {author} {\bibfnamefont {M.~R.}\ \bibnamefont
  {Setare}}, \ and\ \bibinfo {author} {\bibfnamefont {S.}~\bibnamefont
  {Zarepour}},\ }\href {\doibase 10.1140/epjp/s13360-022-03066-y} {\bibfield
  {journal} {\bibinfo  {journal} {Eur. Phys. J. Plus}\ }\textbf {\bibinfo
  {volume} {137}},\ \bibinfo {pages} {859} (\bibinfo {year} {2022})},\ \bibinfo
  {note} {[Erratum: Eur.Phys.J.Plus 137, 1090 (2022)]},\ \Eprint
  {http://arxiv.org/abs/2112.03757} {arXiv:2112.03757 [hep-th]} \BibitemShut
  {NoStop}%
\bibitem [{\citenamefont {Gibbons}(2001)}]{Gibbons:2001gy}%
  \BibitemOpen
  \bibfield  {author} {\bibinfo {author} {\bibfnamefont {G.~W.}\ \bibnamefont
  {Gibbons}},\ }\href {\doibase 10.1063/1.1419338} {\bibfield  {journal}
  {\bibinfo  {journal} {AIP Conf. Proc.}\ }\textbf {\bibinfo {volume} {589}},\
  \bibinfo {pages} {324} (\bibinfo {year} {2001})},\ \Eprint
  {http://arxiv.org/abs/hep-th/0106059} {arXiv:hep-th/0106059} \BibitemShut
  {NoStop}%
\bibitem [{\citenamefont {Bandos}\ \emph {et~al.}(2021)\citenamefont {Bandos},
  \citenamefont {Lechner}, \citenamefont {Sorokin},\ and\ \citenamefont
  {Townsend}}]{Bandos:2020hgy}%
  \BibitemOpen
  \bibfield  {author} {\bibinfo {author} {\bibfnamefont {I.}~\bibnamefont
  {Bandos}}, \bibinfo {author} {\bibfnamefont {K.}~\bibnamefont {Lechner}},
  \bibinfo {author} {\bibfnamefont {D.}~\bibnamefont {Sorokin}}, \ and\
  \bibinfo {author} {\bibfnamefont {P.~K.}\ \bibnamefont {Townsend}},\ }\href
  {\doibase 10.1007/JHEP03(2021)022} {\bibfield  {journal} {\bibinfo  {journal}
  {JHEP}\ }\textbf {\bibinfo {volume} {03}},\ \bibinfo {pages} {022} (\bibinfo
  {year} {2021})},\ \Eprint {http://arxiv.org/abs/2012.09286} {arXiv:2012.09286
  [hep-th]} \BibitemShut {NoStop}%
\bibitem [{\citenamefont {Aaboud}\ \emph {et~al.}(2017)\citenamefont {Aaboud}
  \emph {et~al.}}]{ATLAS:2017fur}%
  \BibitemOpen
  \bibfield  {author} {\bibinfo {author} {\bibfnamefont {M.}~\bibnamefont
  {Aaboud}} \emph {et~al.} (\bibinfo {collaboration} {ATLAS}),\ }\href
  {\doibase 10.1038/nphys4208} {\bibfield  {journal} {\bibinfo  {journal}
  {Nature Phys.}\ }\textbf {\bibinfo {volume} {13}},\ \bibinfo {pages} {852}
  (\bibinfo {year} {2017})},\ \Eprint {http://arxiv.org/abs/1702.01625}
  {arXiv:1702.01625 [hep-ex]} \BibitemShut {NoStop}%
\bibitem [{\citenamefont {Akhmadaliev}\ \emph {et~al.}(2002)\citenamefont
  {Akhmadaliev} \emph {et~al.}}]{Akhmadaliev:2001ik}%
  \BibitemOpen
  \bibfield  {author} {\bibinfo {author} {\bibfnamefont {S.~Z.}\ \bibnamefont
  {Akhmadaliev}} \emph {et~al.},\ }\href {\doibase
  10.1103/PhysRevLett.89.061802} {\bibfield  {journal} {\bibinfo  {journal}
  {Phys. Rev. Lett.}\ }\textbf {\bibinfo {volume} {89}},\ \bibinfo {pages}
  {061802} (\bibinfo {year} {2002})},\ \Eprint
  {http://arxiv.org/abs/hep-ex/0111084} {arXiv:hep-ex/0111084} \BibitemShut
  {NoStop}%
\bibitem [{\citenamefont {Luiten}\ and\ \citenamefont
  {Petersen}(2004)}]{Luiten:2004py}%
  \BibitemOpen
  \bibfield  {author} {\bibinfo {author} {\bibfnamefont {A.~N.}\ \bibnamefont
  {Luiten}}\ and\ \bibinfo {author} {\bibfnamefont {J.~C.}\ \bibnamefont
  {Petersen}},\ }\href {\doibase 10.1016/j.physleta.2004.08.020} {\bibfield
  {journal} {\bibinfo  {journal} {Phys. Lett. A}\ }\textbf {\bibinfo {volume}
  {330}},\ \bibinfo {pages} {429} (\bibinfo {year} {2004})},\ \Eprint
  {http://arxiv.org/abs/physics/0402071} {arXiv:physics/0402071} \BibitemShut
  {NoStop}%
\bibitem [{\citenamefont {Brodin}\ \emph {et~al.}(2001)\citenamefont {Brodin},
  \citenamefont {Marklund},\ and\ \citenamefont {Stenflo}}]{Brodin:2001zz}%
  \BibitemOpen
  \bibfield  {author} {\bibinfo {author} {\bibfnamefont {G.}~\bibnamefont
  {Brodin}}, \bibinfo {author} {\bibfnamefont {M.}~\bibnamefont {Marklund}}, \
  and\ \bibinfo {author} {\bibfnamefont {L.}~\bibnamefont {Stenflo}},\ }\href
  {\doibase 10.1103/PhysRevLett.87.171801} {\bibfield  {journal} {\bibinfo
  {journal} {Phys. Rev. Lett.}\ }\textbf {\bibinfo {volume} {87}},\ \bibinfo
  {pages} {171801} (\bibinfo {year} {2001})},\ \Eprint
  {http://arxiv.org/abs/physics/0108022} {arXiv:physics/0108022} \BibitemShut
  {NoStop}%
\bibitem [{\citenamefont {Bolmont}\ and\ \citenamefont
  {Perennes}(2020)}]{Bolmont:2020aed}%
  \BibitemOpen
  \bibfield  {author} {\bibinfo {author} {\bibfnamefont {J.}~\bibnamefont
  {Bolmont}}\ and\ \bibinfo {author} {\bibfnamefont {C.}~\bibnamefont
  {Perennes}},\ }\href {\doibase 10.1088/1742-6596/1586/1/012033} {\bibfield
  {journal} {\bibinfo  {journal} {J. Phys. Conf. Ser.}\ }\textbf {\bibinfo
  {volume} {1586}},\ \bibinfo {pages} {012033} (\bibinfo {year}
  {2020})}\BibitemShut {NoStop}%
\bibitem [{\citenamefont {Bandos}\ \emph {et~al.}(2020)\citenamefont {Bandos},
  \citenamefont {Lechner}, \citenamefont {Sorokin},\ and\ \citenamefont
  {Townsend}}]{Bandos:2020jsw}%
  \BibitemOpen
  \bibfield  {author} {\bibinfo {author} {\bibfnamefont {I.}~\bibnamefont
  {Bandos}}, \bibinfo {author} {\bibfnamefont {K.}~\bibnamefont {Lechner}},
  \bibinfo {author} {\bibfnamefont {D.}~\bibnamefont {Sorokin}}, \ and\
  \bibinfo {author} {\bibfnamefont {P.~K.}\ \bibnamefont {Townsend}},\ }\href
  {\doibase 10.1103/PhysRevD.102.121703} {\bibfield  {journal} {\bibinfo
  {journal} {Phys. Rev. D}\ }\textbf {\bibinfo {volume} {102}},\ \bibinfo
  {pages} {121703} (\bibinfo {year} {2020})},\ \Eprint
  {http://arxiv.org/abs/2007.09092} {arXiv:2007.09092 [hep-th]} \BibitemShut
  {NoStop}%
\bibitem [{\citenamefont {Neves}\ \emph {et~al.}(2021)\citenamefont {Neves},
  \citenamefont {de~Oliveira}, \citenamefont {Ospedal},\ and\ \citenamefont
  {Helay\"el-Neto}}]{Neves:2021tbt}%
  \BibitemOpen
  \bibfield  {author} {\bibinfo {author} {\bibfnamefont {M.~J.}\ \bibnamefont
  {Neves}}, \bibinfo {author} {\bibfnamefont {J.~B.}\ \bibnamefont
  {de~Oliveira}}, \bibinfo {author} {\bibfnamefont {L.~P.~R.}\ \bibnamefont
  {Ospedal}}, \ and\ \bibinfo {author} {\bibfnamefont {J.~A.}\ \bibnamefont
  {Helay\"el-Neto}},\ }\href {\doibase 10.1103/PhysRevD.104.015006} {\bibfield
  {journal} {\bibinfo  {journal} {Phys. Rev. D}\ }\textbf {\bibinfo {volume}
  {104}},\ \bibinfo {pages} {015006} (\bibinfo {year} {2021})},\ \Eprint
  {http://arxiv.org/abs/2101.03642} {arXiv:2101.03642 [hep-th]} \BibitemShut
  {NoStop}%
\bibitem [{\citenamefont {Novello}\ \emph {et~al.}(2000)\citenamefont
  {Novello}, \citenamefont {De~Lorenci}, \citenamefont {Salim},\ and\
  \citenamefont {Klippert}}]{Novello:1999pg}%
  \BibitemOpen
  \bibfield  {author} {\bibinfo {author} {\bibfnamefont {M.}~\bibnamefont
  {Novello}}, \bibinfo {author} {\bibfnamefont {V.~A.}\ \bibnamefont
  {De~Lorenci}}, \bibinfo {author} {\bibfnamefont {J.~M.}\ \bibnamefont
  {Salim}}, \ and\ \bibinfo {author} {\bibfnamefont {R.}~\bibnamefont
  {Klippert}},\ }\href {\doibase 10.1103/PhysRevD.61.045001} {\bibfield
  {journal} {\bibinfo  {journal} {Phys. Rev. D}\ }\textbf {\bibinfo {volume}
  {61}},\ \bibinfo {pages} {045001} (\bibinfo {year} {2000})},\ \Eprint
  {http://arxiv.org/abs/gr-qc/9911085} {arXiv:gr-qc/9911085} \BibitemShut
  {NoStop}%
\bibitem [{\citenamefont {de~Melo}\ \emph {et~al.}(2015)\citenamefont
  {de~Melo}, \citenamefont {Medeiros},\ and\ \citenamefont
  {Pompeia}}]{deMelo:2014isa}%
  \BibitemOpen
  \bibfield  {author} {\bibinfo {author} {\bibfnamefont {C.~A.~M.}\
  \bibnamefont {de~Melo}}, \bibinfo {author} {\bibfnamefont {L.~G.}\
  \bibnamefont {Medeiros}}, \ and\ \bibinfo {author} {\bibfnamefont {P.~J.}\
  \bibnamefont {Pompeia}},\ }\href {\doibase 10.1142/S021773231550025X}
  {\bibfield  {journal} {\bibinfo  {journal} {Mod. Phys. Lett. A}\ }\textbf
  {\bibinfo {volume} {30}},\ \bibinfo {pages} {1550025} (\bibinfo {year}
  {2015})},\ \Eprint {http://arxiv.org/abs/1407.0567} {arXiv:1407.0567
  [hep-th]} \BibitemShut {NoStop}%
\bibitem [{\citenamefont {Liberati}\ \emph {et~al.}(2001)\citenamefont
  {Liberati}, \citenamefont {Sonego},\ and\ \citenamefont
  {Visser}}]{Liberati:2000mp}%
  \BibitemOpen
  \bibfield  {author} {\bibinfo {author} {\bibfnamefont {S.}~\bibnamefont
  {Liberati}}, \bibinfo {author} {\bibfnamefont {S.}~\bibnamefont {Sonego}}, \
  and\ \bibinfo {author} {\bibfnamefont {M.}~\bibnamefont {Visser}},\ }\href
  {\doibase 10.1103/PhysRevD.63.085003} {\bibfield  {journal} {\bibinfo
  {journal} {Phys. Rev. D}\ }\textbf {\bibinfo {volume} {63}},\ \bibinfo
  {pages} {085003} (\bibinfo {year} {2001})},\ \Eprint
  {http://arxiv.org/abs/quant-ph/0010055} {arXiv:quant-ph/0010055} \BibitemShut
  {NoStop}%
\bibitem [{\citenamefont {Barton}(1991)}]{Barton:1990mu}%
  \BibitemOpen
  \bibfield  {author} {\bibinfo {author} {\bibfnamefont {G.}~\bibnamefont
  {Barton}},\ }\href {\doibase 10.1016/0003-4916(91)90237-3} {\bibfield
  {journal} {\bibinfo  {journal} {Annals Phys.}\ }\textbf {\bibinfo {volume}
  {205}},\ \bibinfo {pages} {49} (\bibinfo {year} {1991})}\BibitemShut
  {NoStop}%
\bibitem [{\citenamefont {Kong}\ and\ \citenamefont
  {Ravndal}(1998)}]{Kong:1998ic}%
  \BibitemOpen
  \bibfield  {author} {\bibinfo {author} {\bibfnamefont {X.-w.}\ \bibnamefont
  {Kong}}\ and\ \bibinfo {author} {\bibfnamefont {F.}~\bibnamefont {Ravndal}},\
  }\href {\doibase 10.1016/S0550-3213(98)00364-2} {\bibfield  {journal}
  {\bibinfo  {journal} {Nucl. Phys. B}\ }\textbf {\bibinfo {volume} {526}},\
  \bibinfo {pages} {627} (\bibinfo {year} {1998})},\ \Eprint
  {http://arxiv.org/abs/hep-ph/9803216} {arXiv:hep-ph/9803216} \BibitemShut
  {NoStop}%
\bibitem [{\citenamefont {Niau~Akmansoy}\ and\ \citenamefont
  {Gouv\^ea~Medeiros}(2014)}]{NiauAkmansoy:2013sxs}%
  \BibitemOpen
  \bibfield  {author} {\bibinfo {author} {\bibfnamefont {P.}~\bibnamefont
  {Niau~Akmansoy}}\ and\ \bibinfo {author} {\bibfnamefont {L.}~\bibnamefont
  {Gouv\^ea~Medeiros}},\ }\href {\doibase 10.1016/j.physletb.2014.10.003}
  {\bibfield  {journal} {\bibinfo  {journal} {Phys. Lett. B}\ }\textbf
  {\bibinfo {volume} {738}},\ \bibinfo {pages} {317} (\bibinfo {year}
  {2014})},\ \Eprint {http://arxiv.org/abs/1311.4917} {arXiv:1311.4917
  [hep-ph]} \BibitemShut {NoStop}%
\bibitem [{\citenamefont {Anacleto}\ \emph {et~al.}(2018)\citenamefont
  {Anacleto}, \citenamefont {Brito}, \citenamefont {Maciel}, \citenamefont
  {Mohammadi}, \citenamefont {Passos}, \citenamefont {Santos},\ and\
  \citenamefont {Santos}}]{Anacleto:2018wlj}%
  \BibitemOpen
  \bibfield  {author} {\bibinfo {author} {\bibfnamefont {M.~A.}\ \bibnamefont
  {Anacleto}}, \bibinfo {author} {\bibfnamefont {F.~A.}\ \bibnamefont {Brito}},
  \bibinfo {author} {\bibfnamefont {E.}~\bibnamefont {Maciel}}, \bibinfo
  {author} {\bibfnamefont {A.}~\bibnamefont {Mohammadi}}, \bibinfo {author}
  {\bibfnamefont {E.}~\bibnamefont {Passos}}, \bibinfo {author} {\bibfnamefont
  {W.~O.}\ \bibnamefont {Santos}}, \ and\ \bibinfo {author} {\bibfnamefont
  {J.~R.~L.}\ \bibnamefont {Santos}},\ }\href {\doibase
  10.1016/j.physletb.2018.08.043} {\bibfield  {journal} {\bibinfo  {journal}
  {Phys. Lett. B}\ }\textbf {\bibinfo {volume} {785}},\ \bibinfo {pages} {191}
  (\bibinfo {year} {2018})},\ \Eprint {http://arxiv.org/abs/1806.08273}
  {arXiv:1806.08273 [hep-th]} \BibitemShut {NoStop}%
\bibitem [{\citenamefont {Heisenberg}\ and\ \citenamefont
  {Euler}(1936)}]{Heisenberg:1936nmg}%
  \BibitemOpen
  \bibfield  {author} {\bibinfo {author} {\bibfnamefont {W.}~\bibnamefont
  {Heisenberg}}\ and\ \bibinfo {author} {\bibfnamefont {H.}~\bibnamefont
  {Euler}},\ }\href {\doibase 10.1007/BF01343663} {\bibfield  {journal}
  {\bibinfo  {journal} {Z. Phys.}\ }\textbf {\bibinfo {volume} {98}},\ \bibinfo
  {pages} {714} (\bibinfo {year} {1936})},\ \Eprint
  {http://arxiv.org/abs/physics/0605038} {arXiv:physics/0605038} \BibitemShut
  {NoStop}%
\bibitem [{\citenamefont {Dunne}(2012)}]{Dunne:2012vv}%
  \BibitemOpen
  \bibfield  {author} {\bibinfo {author} {\bibfnamefont {G.~V.}\ \bibnamefont
  {Dunne}},\ }\href {\doibase 10.1142/S0217751X12600044} {\bibfield  {journal}
  {\bibinfo  {journal} {Int. J. Mod. Phys. A}\ }\textbf {\bibinfo {volume}
  {27}},\ \bibinfo {pages} {1260004} (\bibinfo {year} {2012})},\ \Eprint
  {http://arxiv.org/abs/1202.1557} {arXiv:1202.1557 [hep-th]} \BibitemShut
  {NoStop}%
\bibitem [{\citenamefont {Adler}\ \emph {et~al.}(1970)\citenamefont {Adler},
  \citenamefont {Bahcall}, \citenamefont {Callan},\ and\ \citenamefont
  {Rosenbluth}}]{Adler:1970gg}%
  \BibitemOpen
  \bibfield  {author} {\bibinfo {author} {\bibfnamefont {S.~L.}\ \bibnamefont
  {Adler}}, \bibinfo {author} {\bibfnamefont {J.~N.}\ \bibnamefont {Bahcall}},
  \bibinfo {author} {\bibfnamefont {C.~G.}\ \bibnamefont {Callan}}, \ and\
  \bibinfo {author} {\bibfnamefont {M.~N.}\ \bibnamefont {Rosenbluth}},\ }\href
  {\doibase 10.1103/PhysRevLett.25.1061} {\bibfield  {journal} {\bibinfo
  {journal} {Phys. Rev. Lett.}\ }\textbf {\bibinfo {volume} {25}},\ \bibinfo
  {pages} {1061} (\bibinfo {year} {1970})}\BibitemShut {NoStop}%
\bibitem [{\citenamefont {Ritus}(1972)}]{Ritus:1972ky}%
  \BibitemOpen
  \bibfield  {author} {\bibinfo {author} {\bibfnamefont {V.~I.}\ \bibnamefont
  {Ritus}},\ }\href {\doibase 10.1016/0003-4916(72)90191-1} {\bibfield
  {journal} {\bibinfo  {journal} {Annals Phys.}\ }\textbf {\bibinfo {volume}
  {69}},\ \bibinfo {pages} {555} (\bibinfo {year} {1972})}\BibitemShut
  {NoStop}%
\bibitem [{\citenamefont {Gold}(1968)}]{Gold:1968zf}%
  \BibitemOpen
  \bibfield  {author} {\bibinfo {author} {\bibfnamefont {T.}~\bibnamefont
  {Gold}},\ }\href {\doibase 10.1038/218731a0} {\bibfield  {journal} {\bibinfo
  {journal} {Nature}\ }\textbf {\bibinfo {volume} {218}},\ \bibinfo {pages}
  {731} (\bibinfo {year} {1968})}\BibitemShut {NoStop}%
\bibitem [{\citenamefont {Baring}(2008)}]{Baring:2008aw}%
  \BibitemOpen
  \bibfield  {author} {\bibinfo {author} {\bibfnamefont {M.~G.}\ \bibnamefont
  {Baring}},\ }\href {\doibase 10.1063/1.3020681} {\bibfield  {journal}
  {\bibinfo  {journal} {AIP Conf. Proc.}\ }\textbf {\bibinfo {volume} {1051}},\
  \bibinfo {pages} {53} (\bibinfo {year} {2008})},\ \Eprint
  {http://arxiv.org/abs/0804.0832} {arXiv:0804.0832 [astro-ph]} \BibitemShut
  {NoStop}%
\bibitem [{\citenamefont {Born}\ and\ \citenamefont
  {Infeld}(1934)}]{Born:1934gh}%
  \BibitemOpen
  \bibfield  {author} {\bibinfo {author} {\bibfnamefont {M.}~\bibnamefont
  {Born}}\ and\ \bibinfo {author} {\bibfnamefont {L.}~\bibnamefont {Infeld}},\
  }\href {\doibase 10.1098/rspa.1934.0059} {\bibfield  {journal} {\bibinfo
  {journal} {Proc. Roy. Soc. Lond. A}\ }\textbf {\bibinfo {volume} {144}},\
  \bibinfo {pages} {425} (\bibinfo {year} {1934})}\BibitemShut {NoStop}%
\bibitem [{\citenamefont {Fradkin}\ and\ \citenamefont
  {Tseytlin}(1985)}]{Fradkin:1985qd}%
  \BibitemOpen
  \bibfield  {author} {\bibinfo {author} {\bibfnamefont {E.~S.}\ \bibnamefont
  {Fradkin}}\ and\ \bibinfo {author} {\bibfnamefont {A.~A.}\ \bibnamefont
  {Tseytlin}},\ }\href {\doibase 10.1016/0370-2693(85)90205-9} {\bibfield
  {journal} {\bibinfo  {journal} {Phys. Lett. B}\ }\textbf {\bibinfo {volume}
  {163}},\ \bibinfo {pages} {123} (\bibinfo {year} {1985})}\BibitemShut
  {NoStop}%
\bibitem [{\citenamefont {Bergshoeff}\ \emph {et~al.}(1987)\citenamefont
  {Bergshoeff}, \citenamefont {Sezgin}, \citenamefont {Pope},\ and\
  \citenamefont {Townsend}}]{Bergshoeff:1987at}%
  \BibitemOpen
  \bibfield  {author} {\bibinfo {author} {\bibfnamefont {E.}~\bibnamefont
  {Bergshoeff}}, \bibinfo {author} {\bibfnamefont {E.}~\bibnamefont {Sezgin}},
  \bibinfo {author} {\bibfnamefont {C.~N.}\ \bibnamefont {Pope}}, \ and\
  \bibinfo {author} {\bibfnamefont {P.~K.}\ \bibnamefont {Townsend}},\ }\href
  {\doibase 10.1016/0370-2693(87)90707-6} {\bibfield  {journal} {\bibinfo
  {journal} {Phys. Lett. B}\ }\textbf {\bibinfo {volume} {188}},\ \bibinfo
  {pages} {70} (\bibinfo {year} {1987})}\BibitemShut {NoStop}%
\bibitem [{\citenamefont {Banerjee}\ and\ \citenamefont
  {Roychowdhury}(2012)}]{Banerjee:2012zm}%
  \BibitemOpen
  \bibfield  {author} {\bibinfo {author} {\bibfnamefont {R.}~\bibnamefont
  {Banerjee}}\ and\ \bibinfo {author} {\bibfnamefont {D.}~\bibnamefont
  {Roychowdhury}},\ }\href {\doibase 10.1103/PhysRevD.85.104043} {\bibfield
  {journal} {\bibinfo  {journal} {Phys. Rev. D}\ }\textbf {\bibinfo {volume}
  {85}},\ \bibinfo {pages} {104043} (\bibinfo {year} {2012})},\ \Eprint
  {http://arxiv.org/abs/1203.0118} {arXiv:1203.0118 [gr-qc]} \BibitemShut
  {NoStop}%
\bibitem [{\citenamefont {Gunasekaran}\ \emph {et~al.}(2012)\citenamefont
  {Gunasekaran}, \citenamefont {Mann},\ and\ \citenamefont
  {Kubiznak}}]{Gunasekaran:2012dq}%
  \BibitemOpen
  \bibfield  {author} {\bibinfo {author} {\bibfnamefont {S.}~\bibnamefont
  {Gunasekaran}}, \bibinfo {author} {\bibfnamefont {R.~B.}\ \bibnamefont
  {Mann}}, \ and\ \bibinfo {author} {\bibfnamefont {D.}~\bibnamefont
  {Kubiznak}},\ }\href {\doibase 10.1007/JHEP11(2012)110} {\bibfield  {journal}
  {\bibinfo  {journal} {JHEP}\ }\textbf {\bibinfo {volume} {11}},\ \bibinfo
  {pages} {110} (\bibinfo {year} {2012})},\ \Eprint
  {http://arxiv.org/abs/1208.6251} {arXiv:1208.6251 [hep-th]} \BibitemShut
  {NoStop}%
\bibitem [{\citenamefont {Niau~Akmansoy}\ and\ \citenamefont
  {Medeiros}(2018)}]{NiauAkmansoy:2017kbw}%
  \BibitemOpen
  \bibfield  {author} {\bibinfo {author} {\bibfnamefont {P.}~\bibnamefont
  {Niau~Akmansoy}}\ and\ \bibinfo {author} {\bibfnamefont {L.~G.}\ \bibnamefont
  {Medeiros}},\ }\href {\doibase 10.1140/epjc/s10052-018-5643-1} {\bibfield
  {journal} {\bibinfo  {journal} {Eur. Phys. J. C}\ }\textbf {\bibinfo {volume}
  {78}},\ \bibinfo {pages} {143} (\bibinfo {year} {2018})},\ \Eprint
  {http://arxiv.org/abs/1712.05486} {arXiv:1712.05486 [hep-ph]} \BibitemShut
  {NoStop}%
\bibitem [{\citenamefont {Niau~Akmansoy}\ and\ \citenamefont
  {Medeiros}(2019)}]{NiauAkmansoy:2018ilv}%
  \BibitemOpen
  \bibfield  {author} {\bibinfo {author} {\bibfnamefont {P.}~\bibnamefont
  {Niau~Akmansoy}}\ and\ \bibinfo {author} {\bibfnamefont {L.~G.}\ \bibnamefont
  {Medeiros}},\ }\href {\doibase 10.1103/PhysRevD.99.115005} {\bibfield
  {journal} {\bibinfo  {journal} {Phys. Rev. D}\ }\textbf {\bibinfo {volume}
  {99}},\ \bibinfo {pages} {115005} (\bibinfo {year} {2019})},\ \Eprint
  {http://arxiv.org/abs/1809.01296} {arXiv:1809.01296 [hep-ph]} \BibitemShut
  {NoStop}%
\bibitem [{\citenamefont {Kruglov}(2010)}]{Kruglov:2009he}%
  \BibitemOpen
  \bibfield  {author} {\bibinfo {author} {\bibfnamefont {S.~I.}\ \bibnamefont
  {Kruglov}},\ }\href {\doibase 10.1088/1751-8113/43/37/375402} {\bibfield
  {journal} {\bibinfo  {journal} {J. Phys. A}\ }\textbf {\bibinfo {volume}
  {43}},\ \bibinfo {pages} {375402} (\bibinfo {year} {2010})},\ \Eprint
  {http://arxiv.org/abs/0909.1032} {arXiv:0909.1032 [hep-th]} \BibitemShut
  {NoStop}%
\bibitem [{\citenamefont {Braaten}\ and\ \citenamefont
  {Nieto}(1995)}]{Braaten:1995cm}%
  \BibitemOpen
  \bibfield  {author} {\bibinfo {author} {\bibfnamefont {E.}~\bibnamefont
  {Braaten}}\ and\ \bibinfo {author} {\bibfnamefont {A.}~\bibnamefont
  {Nieto}},\ }\href {\doibase 10.1103/PhysRevD.51.6990} {\bibfield  {journal}
  {\bibinfo  {journal} {Phys. Rev. D}\ }\textbf {\bibinfo {volume} {51}},\
  \bibinfo {pages} {6990} (\bibinfo {year} {1995})},\ \Eprint
  {http://arxiv.org/abs/hep-ph/9501375} {arXiv:hep-ph/9501375} \BibitemShut
  {NoStop}%
\bibitem [{\citenamefont {Ramos}\ and\ \citenamefont
  {Filho}(2014)}]{Ramos:2014mda}%
  \BibitemOpen
  \bibfield  {author} {\bibinfo {author} {\bibfnamefont {R.}~\bibnamefont
  {Ramos}}\ and\ \bibinfo {author} {\bibfnamefont {H.~B.}\ \bibnamefont
  {Filho}},\ }\href@noop {} {\  (\bibinfo {year} {2014})},\ \Eprint
  {http://arxiv.org/abs/1409.7374} {arXiv:1409.7374 [hep-th]} \BibitemShut
  {NoStop}%
\bibitem [{\citenamefont {Franklin}\ and\ \citenamefont
  {Garon}(2011)}]{Franklin:2011zz}%
  \BibitemOpen
  \bibfield  {author} {\bibinfo {author} {\bibfnamefont {J.}~\bibnamefont
  {Franklin}}\ and\ \bibinfo {author} {\bibfnamefont {T.}~\bibnamefont
  {Garon}},\ }\href {\doibase 10.1016/j.physleta.2011.02.012} {\bibfield
  {journal} {\bibinfo  {journal} {Phys. Lett. A}\ }\textbf {\bibinfo {volume}
  {375}},\ \bibinfo {pages} {1391} (\bibinfo {year} {2011})},\ \Eprint
  {http://arxiv.org/abs/1102.2277} {arXiv:1102.2277 [quant-ph]} \BibitemShut
  {NoStop}%
\bibitem [{\citenamefont {Ellis}\ \emph {et~al.}(2017)\citenamefont {Ellis},
  \citenamefont {Mavromatos},\ and\ \citenamefont {You}}]{Ellis:2017edi}%
  \BibitemOpen
  \bibfield  {author} {\bibinfo {author} {\bibfnamefont {J.}~\bibnamefont
  {Ellis}}, \bibinfo {author} {\bibfnamefont {N.~E.}\ \bibnamefont
  {Mavromatos}}, \ and\ \bibinfo {author} {\bibfnamefont {T.}~\bibnamefont
  {You}},\ }\href {\doibase 10.1103/PhysRevLett.118.261802} {\bibfield
  {journal} {\bibinfo  {journal} {Phys. Rev. Lett.}\ }\textbf {\bibinfo
  {volume} {118}},\ \bibinfo {pages} {261802} (\bibinfo {year} {2017})},\
  \Eprint {http://arxiv.org/abs/1703.08450} {arXiv:1703.08450 [hep-ph]}
  \BibitemShut {NoStop}%
\end{thebibliography}%

\end{document}